\documentclass[%
 reprint,
superscriptaddress,
 amsmath,amssymb,
 aps,
prb,
]{revtex4-2}

\usepackage{graphicx}% Include figure files
\usepackage{dcolumn}% Align table columns on decimal point
\usepackage{bm}% bold math
\usepackage{braket}
\usepackage{float}
\usepackage{amsmath}
\usepackage{amsfonts}   % if you want the fonts
\usepackage{amssymb}    % if you want extra symbols
\usepackage{comment}

\usepackage{array}
\newcolumntype{P}[1]{>{\centering\arraybackslash}p{#1}}

\usepackage{booktabs}% publication quality tables 

\begin{document}

\preprint{APS/123-QED}

\title{Anisotropic Magnetism in Gd\textsubscript{2}B\textsubscript{5}}% Force line breaks with \\

\author{Maximilien F. Debbas}
%\email{mdebbas@mit.edu}
\affiliation{Department of Nuclear Science and Engineering, Massachusetts Institute of Technology, Cambridge, MA 02139, USA}%Lines break automatically or can be forced with \\

\author{Takehito Suzuki}
\affiliation{Department of Physics, Toho University, Funabashi, Japan}

\author{Alex H. Mayo}
\affiliation{Department of Physics, Massachusetts Institute of Technology, Cambridge, MA 02139, USA}

\author{Mun K. Chan}
\affiliation{National High Magnetic Field Laboratory, Los Alamos National Laboratory, Los Alamos, NM 87545, USA}
 
\author{Joseph G. Checkelsky}
\affiliation{Department of Physics, Massachusetts Institute of Technology, Cambridge, MA 02139, USA}

\date{\today}% It is always \today, today, but any date may be explicitly specified

\begin{abstract}
We report the synthesis of single crystals of Gd\textsubscript{2}B\textsubscript{5} through a ruthenium-gadolinium flux method. The Gd\textsubscript{2}B\textsubscript{5} system is a member of the monoclinic \textit{P2$_1$/c} (No. 14) space group and realizes lattices of gadolinium atoms in the $(1\,0\,0)$ plane. We characterized the sample through orientation-dependent electrical transport, magnetization, magnetic torque, and heat capacity measurements to probe the magnetic anisotropy of the system and map out its phase diagram. Gd\textsubscript{2}B\textsubscript{5} realizes two zero-field ordered phases M\textsubscript{1} and M\textsubscript{2}, as well as a third field-induced ordered phase M$_\perp$ arising when the magnetic field is applied in the $(1\,0\,0)$ plane.

%\begin{description}
%\item[DOI]
%Put DOI information here.
%\end{description}

\end{abstract}

\maketitle

%\tableofcontents

\section{Introduction}
\label{sec:Introduction}

Gadolinium is well known among the lanthanides for forming crystals with quenched crystal electric field (CEF) effects on the local $4f$-electron wavefunction due to the $L=0$ Russell–Saunders coupled orbital angular momentum of the Gd\textsuperscript{3+} ion. As all the angular momentum of the $4f$-electron wavefunction in Gd\textsuperscript{3+} comes from electron spin (such that $J=S=7/2$), the electronic structure is unaffected by any CEF created by charged ions in a ligand field surrounding the gadolinium sites in a crystal (within a first order approximation of the ligand field) such that the free-space, spin-orbit $4f$ electronic structure remains intact on these sites. Removing the anisotropy induced by the CEF can yield interesting magnetic phases such as skyrmion lattices in some centrosymmetric gadolinium systems. Notable examples of such skyrmion lattice materials include %\cite{doi:10.1021/acs.chemrev.0c00297} with notable examples including 
Gd\textsubscript{2}PdSi\textsubscript{3} \cite{doi:10.1126/science.aau0968}, GdRu\textsubscript{2}Ge\textsubscript{2} \cite{Yoshimochi2024}, GdRu\textsubscript{2}Si\textsubscript{2} \cite{Khanh2020}, and Gd\textsubscript{3}Ru\textsubscript{4}Al\textsubscript{12} \cite{hirschberger2019skyrmionphase}. These examples stand in contrast to typical non-centrosymmetric skyrmion materials stabilized by the Dzyaloshinskii-Moriya (DM) interaction \cite{doi:10.1021/acs.chemrev.0c00297,CAMLEY2023100605}.

These skyrmion phases provide a concrete example of a system wherein a field-induced magnetic texture can strongly effect the electronic behavior of the system. Skyrmion phases are associated with a characteristic topological Hall response whereby the skyrmion lattice induces an emergent magnetic field which deflects conduction electrons via an emergent Lorentz force which adds an additional contribution to the Hall response when the system is field-tuned to the skyrmion phase \cite{doi:10.1021/acs.chemrev.0c00297}. Such a topological Hall response is reported for all of the aforementioned gadolinium-based skyrmion materials \cite{doi:10.1126/science.aau0968,Yoshimochi2024,Khanh2020,hirschberger2019skyrmionphase}. These materials all additionally exhibit metamagnetic transitions characterized by kinks or jumps in isothermal magnetization associated with the field-induced magnetic phases (Gd\textsubscript{2}PdSi\textsubscript{3} \cite{doi:10.1021/acs.chemrev.0c00297}, GdRu\textsubscript{2}Ge\textsubscript{2} \cite{Yoshimochi2024,GARNIER199680}, GdRu\textsubscript{2}Si\textsubscript{2} \cite{Khanh2020,GARNIER1995899,GARNIER199680}, and Gd\textsubscript{3}Ru\textsubscript{4}Al\textsubscript{12} \cite{chandragiri2016magnetic}). Such metamagnetic transitions may also be found in other non skyrmion-hosting gadolinium systems such as GdAuGe \cite{PhysRevB.110.064409,PhysRevB.108.235107} and are associated with an anomalous Hall effect proportional to the magnetization across the transition, but no topological Hall effect. 

Gd\textsubscript{2}B\textsubscript{5} is a member of the monoclinic $P2_1/c$ space group that realizes a unique structure; there do not appear to be any other isostructural compounds reported in the literature. A previous structural report exists for Gd\textsubscript{2}B\textsubscript{5} wherein plate-like $0.2 \times 0.2 \times 0.04 \, \text{mm}^3$ single crystals were grown through a GdCl\textsubscript{3} flux growth \cite{schwarz1987kristallstruktur}. This report details the synthesis of large, single crystals of Gd\textsubscript{2}B\textsubscript{5} through a new ruthenium-gadolinium flux growth method. Through orientation-dependent electrical transport, magnetization and heat capacity measurements, a rich and highly anisotropic magnetic phase diagram is uncovered wherein two zero-field ordered phases (M\textsubscript{1} and M\textsubscript{2}) occur, and a third field-induced ordered phase M$_\perp$ arises for certain orientations of the applied magnetic field relative to the crystal. Notably, the onset of the M$_\perp$ phase appears to strongly effect the electronic behavior of the system. The aforementioned growth method moreover yields crystals of sufficient quality to observe quantum oscillations for certain field orientations. Gd\textsubscript{2}B\textsubscript{5} provides a new opportunity to study a rich, anisotropic magnetic phase diagram in a potentially unconventional material ripe for expansion through further exploration.

\section{Growth Method}
\label{sec:Growth Method}

%pg. 88 of Notebook 3, 02.27.2023

Single crystals of Gd\textsubscript{2}B\textsubscript{5} were synthesized through a ruthenium-gadolinium flux which takes advantage of the 917\textsuperscript{o}C eutectic point occurring at a gadolinium to ruthenium ratio of 17:3. The samples were prepared by first preparing a Gd:Ru:B ratio of 17:3:2 under argon in a glovebox and loading it into an alumina Canfield crucible sealed in a quartz tube under high vacuum. The materials used were $>$99.9\% purity, gadolinium chips (Sigma-Aldrich), 99.99\% purity ruthenium powder (Sigma-Aldrich), and 99.9999\% purity, -4 mesh boron powder (Thermo Scientific).

The materials were melted together by first heating the growth to 1150\textsuperscript{o}C over 12 hours and holding at that temperature for 24 hours. After this, the growth was cooled to 950\textsuperscript{o}C over four days and then annealed at that temperature for two days. The growth was then centrifuged directly from the hot furnace using a commercial centrifuge.

The growth resulted in a large number of plate-like crystals in the crystal side of the Canfield crucible. A representative single crystal from this growth is shown n Fig. \ref{fig:Gd2S5_struct}(a). The purity of the crystals was checked using powder x-ray diffraction (XRD) (see Appendix \ref{sec: X-ray Diffraction} for more details).

\section{Crystal Structure}
\label{sec:Crystal Structure}

\begin{figure}[H]
	\centering 
	\includegraphics[width=1.0
    \linewidth]{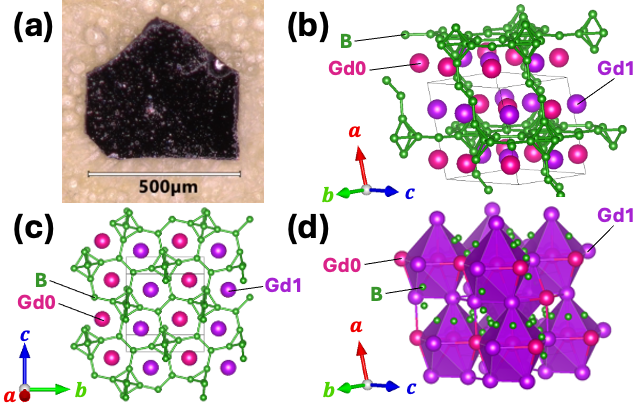}
	\caption{\textbf{(a)} A Gd\textsubscript{2}B\textsubscript{5} single crystal shown with the $(1\,0\,0)$ plane parallel to the page. \textbf{(b)} Gd\textsubscript{2}B\textsubscript{5} crystal structure showing the boron sites in green and two gadolinium Wyckoff sites Gd0 and Gd1 in pink and purple respectively. \textbf{(c)} A view of a boron network layer in the $(1\,0\,0)$ plane. \textbf{(d)} Another view of the Gd\textsubscript{2}B\textsubscript{5} crystal structure showing the network of gadolinium coordination polyhedra. Crystal structures drawn using VESTA \cite{VESTA}}
	\label{fig:Gd2S5_struct}
\end{figure}

The Gd\textsubscript{2}B\textsubscript{5} system is a member of the monoclinic $P2_1/c$ (14) space group. The structure (shown in Figs. \ref{fig:Gd2S5_struct}(b-d)) is a member of the centrosymmetric monoclinic normal (prismatic) crystal class which falls under the $C_{2h}$ point group. The system notably possesses a $2_1$ screw symmetry along the \textit{b} axis and a half-cell glide symmetry along the \textit{c} axis. The crystals all realized a plate-like morphology with the plates growing in the $(1\,0\,0)$ plane (the \textit{bc}-plane). This orientation was determined via both Laue XRD and monochromatic XRD measurements on several single crystals from the same synthesis batch (see Appendix \ref{sec: X-ray Diffraction} for more details).

The boron atoms form network layers parallel to the $(1\,0\,0)$ plane with layers of gadolinium atoms above and below. The boron atoms additionally form ``tower'' structures which cut through the gadolinium layers to connect the boron layers. Note that there are two distinct gadolinium Wyckoff sites (Gd0 and Gd1) in the structure which are drawn in pink and purple respectively in Figs. \ref{fig:Gd2S5_struct}(b-d). The gadolinium atom layers shown in Fig. \ref{fig:Gd2S5_struct}(c) appear to realize a highly distorted square lattice. As shown in Fig. \ref{fig:Gd2S5_struct}(d), the gadolinium structure may also be understood to form a structure of coordination polyhedra, each with a central gadolinium atom that coordinates with eight other gadolinium atoms. This coordination polyhedron forms a highly distorted bicapped trigonal prism (BTP) which leads to an extremely anisotropic magnetic environment around the gadolinium sites.

\section{Measurements}
\label{sec:Measurements}

This section will present results derived from resistivity, heat capacity, magnetization, and magnetic torque measurements on single crystals of Gd\textsubscript{2}B\textsubscript{5} crystals all grown in the same synthesis batch. When referring to crystallographic directions, this section will use the standard notation of \textit{a}, \textit{b}, and \textit{c} for the directions parallel to the $[1\,0\,0]$, $[0\,1\,0]$, and $[0\,0\,1]$ directions respectively. The \textit{a*} axis will refer to the axis perpendicular to the $(1\,0\,0)$ plane. Furthermore, $\theta$ will be used for angles made with the \textit{a*} axis, and $\phi$ will be used for angles in the $(1\,0\,0)$ plane such that the \textit{a*} axis defines a polar coordinate system.

\subsection{Electrical Transport}
\label{Electrical Transport}

Electrical transport measurements were performed in a Quantum Design Physical Property Measurement System (PPMS) on two Gd\textsubscript{2}B\textsubscript{5} devices: device \textit{R1} and device \textit{R2}. Device \textit{R1} was also measured in magnetic fields up to $60 \, \text{T}$ at the National High Magnetic Field Laboratory's Pulsed Field Facility at Los Alamos National Laboratory. Device \textit{R1} was made from a single crystal which had a natural thickness of $25\, \mu\text{m}$ and was affixed to a sapphire substrate with Stycast epoxy. Gold wires were then affixed to the crystal using H20E epoxy, and the crystal was milled with a focused ion beam (FIB) into a six-contact Hall bar device. Device \textit{R2} was made from a single crystal which was manually lapped down to $26\, \mu\text{m}$ with a commercial lapping jig and affixed to a sapphire substrate with GE varnish. Gold wires were then affixed to the crystal using silver paint. In both devices, the current flows in the $(1\,0\,0)$ plane.

\begin{figure}[H]
	\centering 
	\includegraphics[width=0.95
    \linewidth]{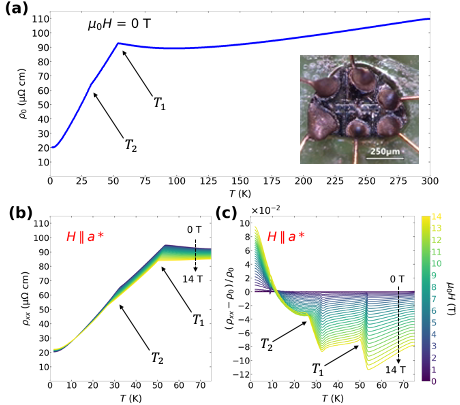}
	\caption{\textbf{(a)} Zero-field longitudinal resistivity of device $R1$ with the two transition temperatures $T_1$ and $T_2$ labeled. The inset shows an image of device $R1$. Longitudinal resistivity \textbf{(b)} and magnetoresistance \textbf{(c)} of device $R1$ in magnetic fields parallel to the \textit{a*} axis.}
	\label{fig:Res00}
\end{figure}

Figure \ref{fig:Res00}(a) shows the zero-field longitudinal resistivity of Gd\textsubscript{2}B\textsubscript{5} as a function of temperature below $300 \, \text{K}$ with two features at temperatures $T_1$ and $T_2$ labeled. The $T_1$ feature appears as a sharp downturn in the resistivity, whereas the $T_2$ feature appears as a kink. These features arise from magnetic phase transitions and manifest themselves clearly in all magnetization measurements of the system. Hereafter, the low temperature magnetic phase existing at $T<T_2$ will be referred to as M\textsubscript{2}, and the intermediate magnetic phase existing at $T_2<T<T_1$ will be referred to as M\textsubscript{1}. The high temperature paramagnetic phase existing at $T>T_1$ will be referred to as PM.

\begin{figure}[H]
	\centering 
	\includegraphics[width=1.0
    \linewidth]{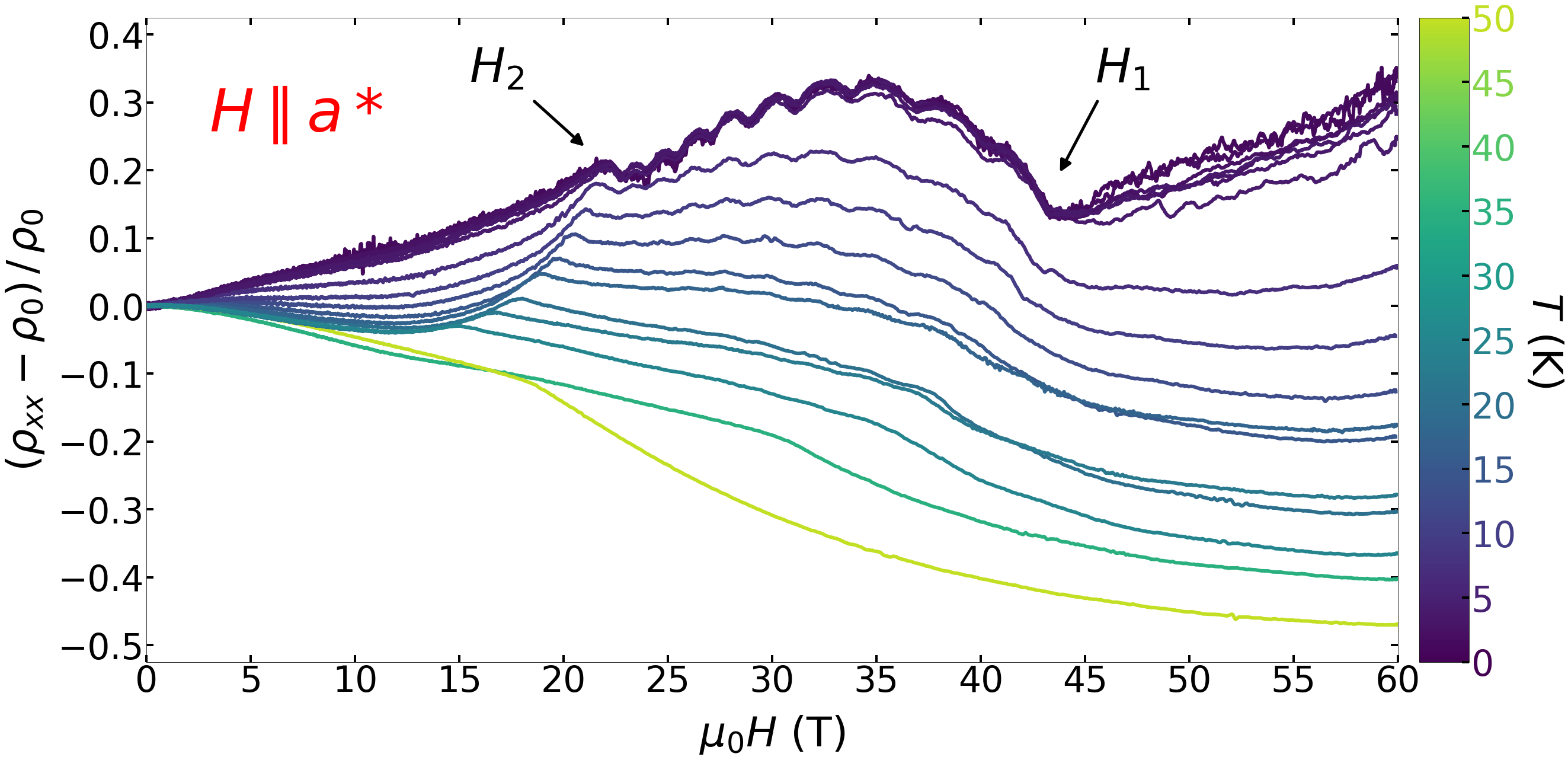}
	\caption{Magnetoresistance of device $R1$ with magnetic field applied  parallel to the \textit{a*} axis. The labels H\textsubscript{1} and H\textsubscript{2} indicate the fields corresponding to the PM$\leftrightarrow$M\textsubscript{1} and M\textsubscript{1}$\leftrightarrow$M\textsubscript{2} phase transitions respectively at $1.8 \, \text{K}$ for this field orientation.}
	\label{fig:HsweepLANL}
\end{figure}

Figure \ref{fig:HsweepLANL} shows the magnetoresistance measured in magnetic fields up to $60 \, \text{T}$ applied parallel to the \textit{a*} axis at Los Alamos National Laboratory. The magnetoresistance exhibits a kink when the M\textsubscript{1}$\leftrightarrow$M\textsubscript{2} phase boundary is crossed, and a kink when the PM$\leftrightarrow$M\textsubscript{1} phase boundary is crossed with prominent quantum oscillations in the M\textsubscript{1} phase. The Hall channel also exhibits a kink when the M\textsubscript{1}$\leftrightarrow$M\textsubscript{2} phase boundary is crossed (see Appendix \ref{sec: Additional Phase Diagram Details} for more details).

\begin{figure}[H]
	\centering 
	\includegraphics[width=1.0
    \linewidth]{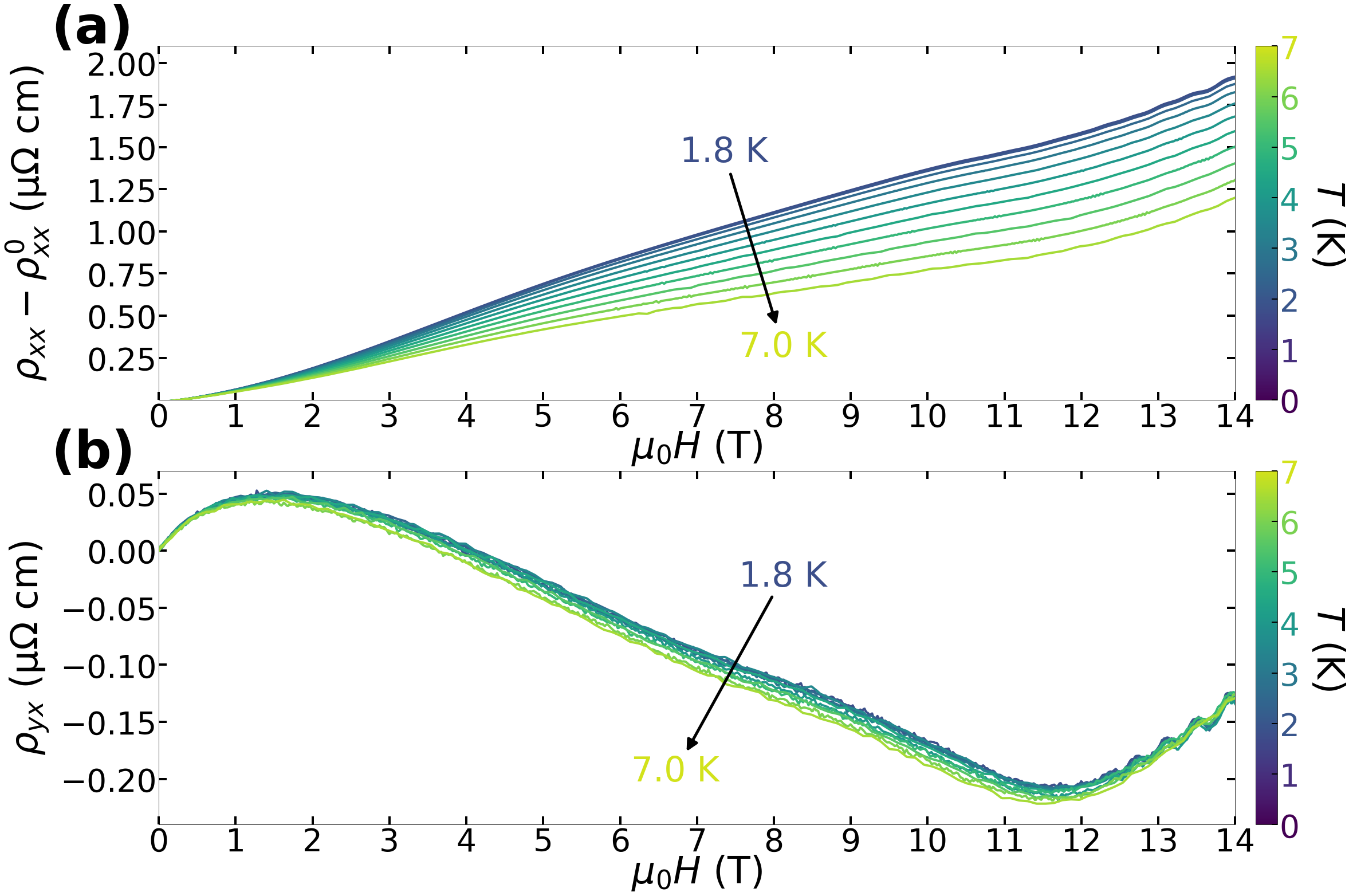}
	\caption{Longitudinal \textbf{(a)} and Hall \textbf{(b)} resistivity of device $R1$ plotted as a function of magnetic field applied parallel to the \textit{a*} axis at fixed temperatures between $1.8 \, \text{K}$ and $7.0 \, \text{K}$.}
	\label{fig:M2_QO}
\end{figure}

Figure \ref{fig:M2_QO} shows the low temperature longitudinal and Hall resistivity data measured for magnetic fields up to $14 \, \text{T}$ applied parallel to the \textit{a*} axis. The Hall resistivity exhibits a prominent maximum near $1.5 \, \text{T}$ and a prominent minimum near $11.5 \, \text{T}$. Both of these features are associated with slight changes in the curvature of the longitudinal resistivity. Both channels also exhibit quantum oscillations in the M\textsubscript{2} phase which appear after the minimum in the Hall resistivity. 

\subsubsection{Quantum Oscillations}

Shubnikov–de Haas (SdH) quantum oscillations were observed in both the M\textsubscript{1} and M\textsubscript{2} phases for field applied parallel to the \textit{a*} axis. The M\textsubscript{1} phase exhibits three Fermi pockets denoted $\alpha_1$, $\beta_1$, and $\gamma_1$. The M\textsubscript{2} phase exhibits two Fermi pockets denoted $\alpha_2$ and $\beta_2$. The frequencies and effective masses (obtained via Lifshitz-Kosevich fits of the oscillation amplitudes) of these pockets are presented in Tables \ref{T: M1QO} and \ref{T: M2QO}. Appendix \ref{sec: SdH Quantum Oscillations} contains additional details on the analysis of the quantum oscilations.

\begin{table}[H]
\caption{\label{T: M1QO} Frequencies (at $2.1 \, \text{K}$) and effective masses $\tilde{m}$ of the Fermi pockets contributing to SdH quantum oscillations in the longitudinal resistivity channel of the M\textsubscript{1} phase}
\centering{
\begin{tabular}{p{2.5cm} P{2.5cm} P{2.5cm}}
\toprule
  Pocket &  freq. (T) &  $\tilde{m} \equiv m^* / m_e$   \\ [0.8ex] 
  \hline
  $\alpha_1$ & 109 & 0.17 \\ [0.6ex] 
  $\beta_1$ & 416 & 0.17 \\ [0.6ex] 
  $\gamma_1$ & 822 & 0.42 \\ [0.6ex] 
\bottomrule
\end{tabular}
}
\end{table}

\begin{table}[H]
\caption{\label{T: M2QO} Frequencies (at $2.0 \, \text{K}$) and effective masses $\tilde{m}$ of the Fermi pockets contributing to SdH quantum oscillations in the longitudinal resistivity channel of the M\textsubscript{2} phase}
\centering{
\begin{tabular}{p{2.5cm} P{2.5cm} P{2.5cm}}
\toprule
  Pocket &  freq. (T) &  $\tilde{m} \equiv m^* / m_e$   \\ [0.8ex] 
  \hline
  $\alpha_2$ & 184 & 0.31 \\ [0.6ex] 
  $\beta_2$ & 443 & 0.26 \\ [0.6ex] 
\bottomrule
\end{tabular}
}
\end{table}

\subsubsection{In-Plane Rotation}
\label{In-Plane Rotation}

\begin{figure}[H]
	\centering 
	\includegraphics[width=1.0
    \linewidth]{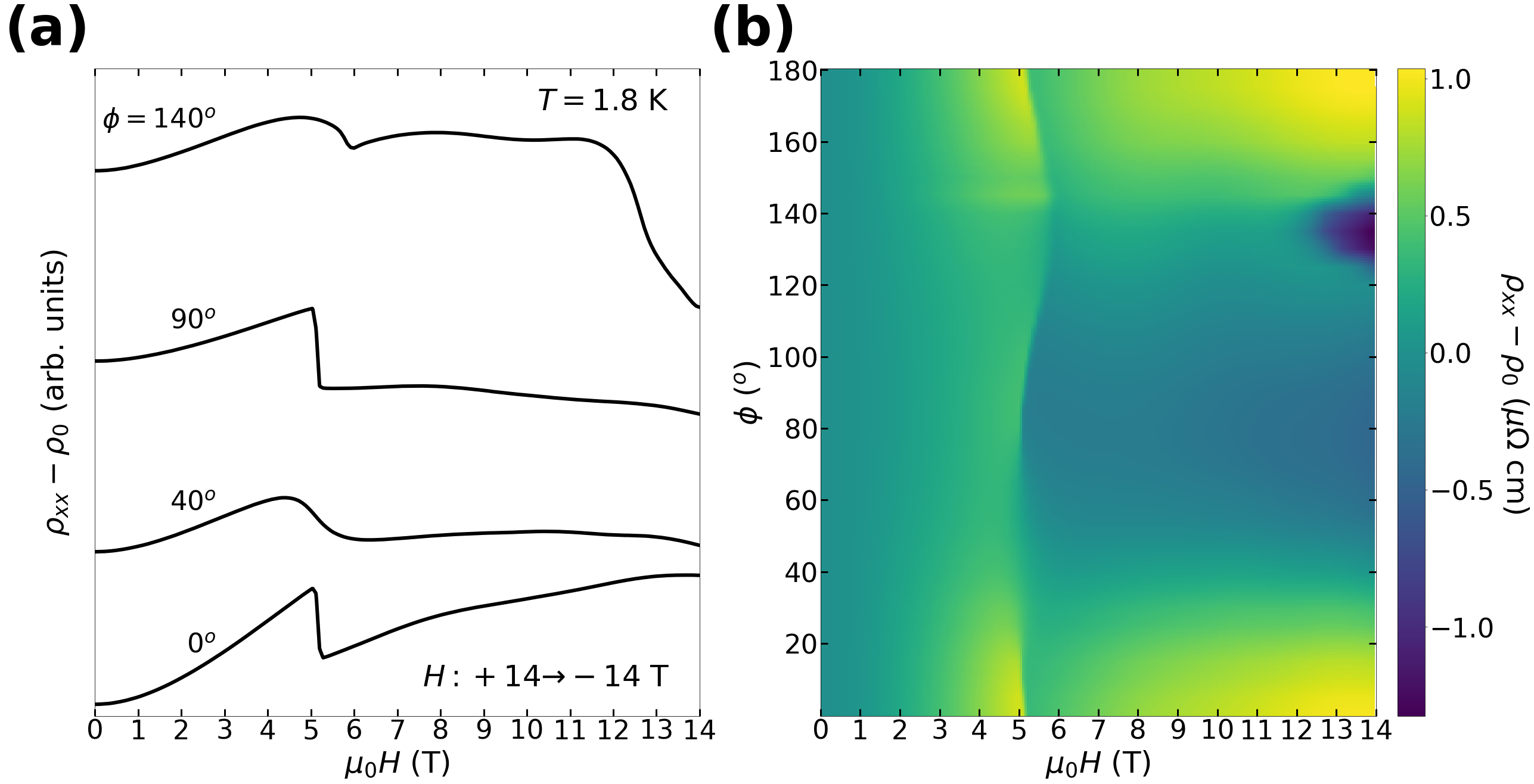}
	\caption{\textbf{(a)} Longitudinal resistivity of device $R1$ at $T = 1.8 \, \text{K}$ plotted as a function of applied field for selected field-angles $\phi$ in the $(1 \, 0 \, 0)$ plane. Note that the magnetic field was swept from $+14 \, \text{T}$ to $-14 \, \text{T}$ and that all of these data have been offset vertically for clarity. \textbf{(b)} A colorplot for the longitudinal resistivity data with applied field on the horizontal axis and in-plane angle on the vertical axis.}
	\label{fig:InPLong}
\end{figure}

\begin{figure}[H]
	\centering 
	\includegraphics[width=1.0
    \linewidth]{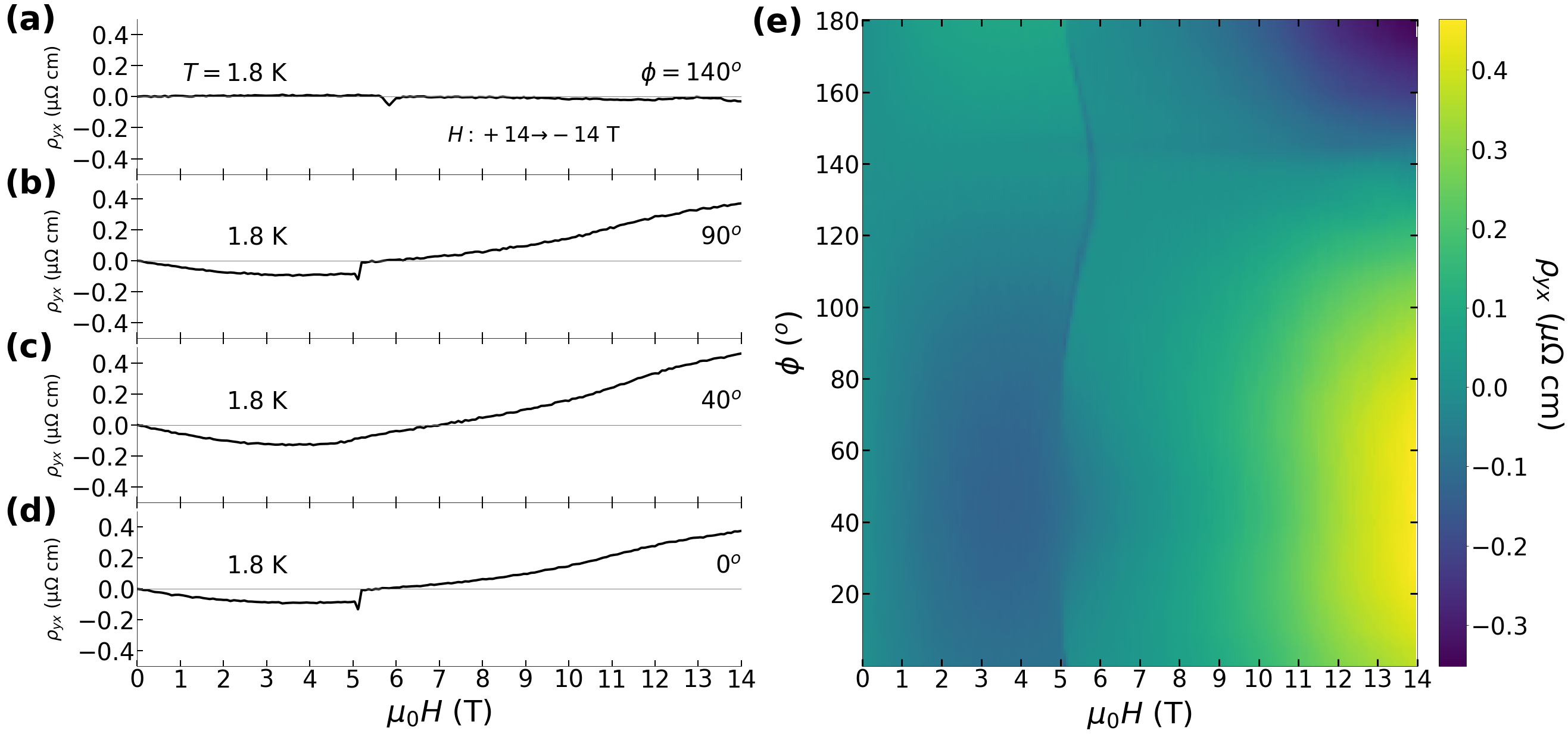}
	\caption{\textbf{(a-d)} Transverse resistivity of device $R1$ at $T = 1.8 \, \text{K}$ plotted as a function of applied field for selected field-angles $\phi$ in the $(1 \, 0 \, 0)$ plane. Note that the magnetic field was swept from $+14 \, \text{T}$ to $-14 \, \text{T}$ and that all of these data have been offset vertically for clarity. \textbf{(e)} A colorplot for the transverse resistivity data with applied field on the horizontal axis and in-plane angle on the vertical axis.}
	\label{fig:InPHall}
\end{figure}

The longitudinal and transverse resistivities were measured at $1.8 \, \text{K}$ for magnetic fields applied perpendicular to the \textit{a*} axis. These measurements were taken at fixed field-angles $\phi$ in the $(1 \, 0 \, 0)$ plane, and colorplots were constructed to show the evolution of the transport data as a function of field-angle. These data are presented in Fig. \ref{fig:InPLong} and Fig. \ref{fig:InPHall} for device $R1$ for field scans taken from $14 \to -14 \, \text{T}$. Note that the device could not be oriented in the $(1 \, 0 \, 0)$ plane, so the absolute values of the field-angle relative to the \textit{b} and \textit{c} axes is arbitrary.

The in-plane magnetic field reveals a prominent feature around $5 \, \text{T}$ which is absent for fields applied parallel to the \textit{a*} axis. This feature is associated with the onset of a field-induced phase that will be referred to as $M_\perp$ as it only occurs when the applied field is close to being in the $(1 \, 0 \, 0)$ plane. An additional feature appears in the longitudinal resistivity channel as a narrow pocket of negative magnetoresistance (at about $\phi = 140$\textsuperscript{o} in Fig. \ref{fig:InPLong} above $12 \, \text{T}$). This feature may signify the termination of the $M_\perp$ for this specific field direction.

Heretofore, all of the electrical transport measurements presented in association with the M$_\perp$ phase have been taken in one direction (sweeping field as $+14 \to -14 \, \text{T}$) and symmetrized (antisymmetrized) to extract the longitudinal (transverse) resistivity for the purposes of making the colorplots in Figs. \ref{fig:InPLong} and \ref{fig:InPHall}. Entering or leaving the M$_\perp$ phase is, however, associated with hysteresis in the electrical transport which necessitates both a downsweep and an upsweep of applied field to properly symmetrize or antisymmetrize the data. Appendix \ref{sec: Hysteresis in Electrical Transport} discusses this hysteretic behavior in greater detail.

\subsection{Magnetization}
\label{Magnetization}

Magnetic susceptibility $\chi$ and magnetization $M$ were measured using SQUID magnetometry in a Quantum Design Magnetic Property Measurement System (MPMS3). Single crystals were affixed to a quartz rod using GE varnish and DC magnetization measurements were performed using Vibrating Sample Magnetometer (VSM) mode. Measurements with field in the $(1 \, 0 \, 0)$ plane were performed on Gd\textsubscript{2}B\textsubscript{5} single crystal which grew thick enough such that it could be visually oriented (as the $\beta$ angle formed between the \textit{a} and \textit{c} axes could clearly be seen). Measurements with field aligned parallel to the \textit{a*} axis were performed on a different single crystal.

\begin{figure}[H]
	\centering 
	\includegraphics[width=1.0
    \linewidth]{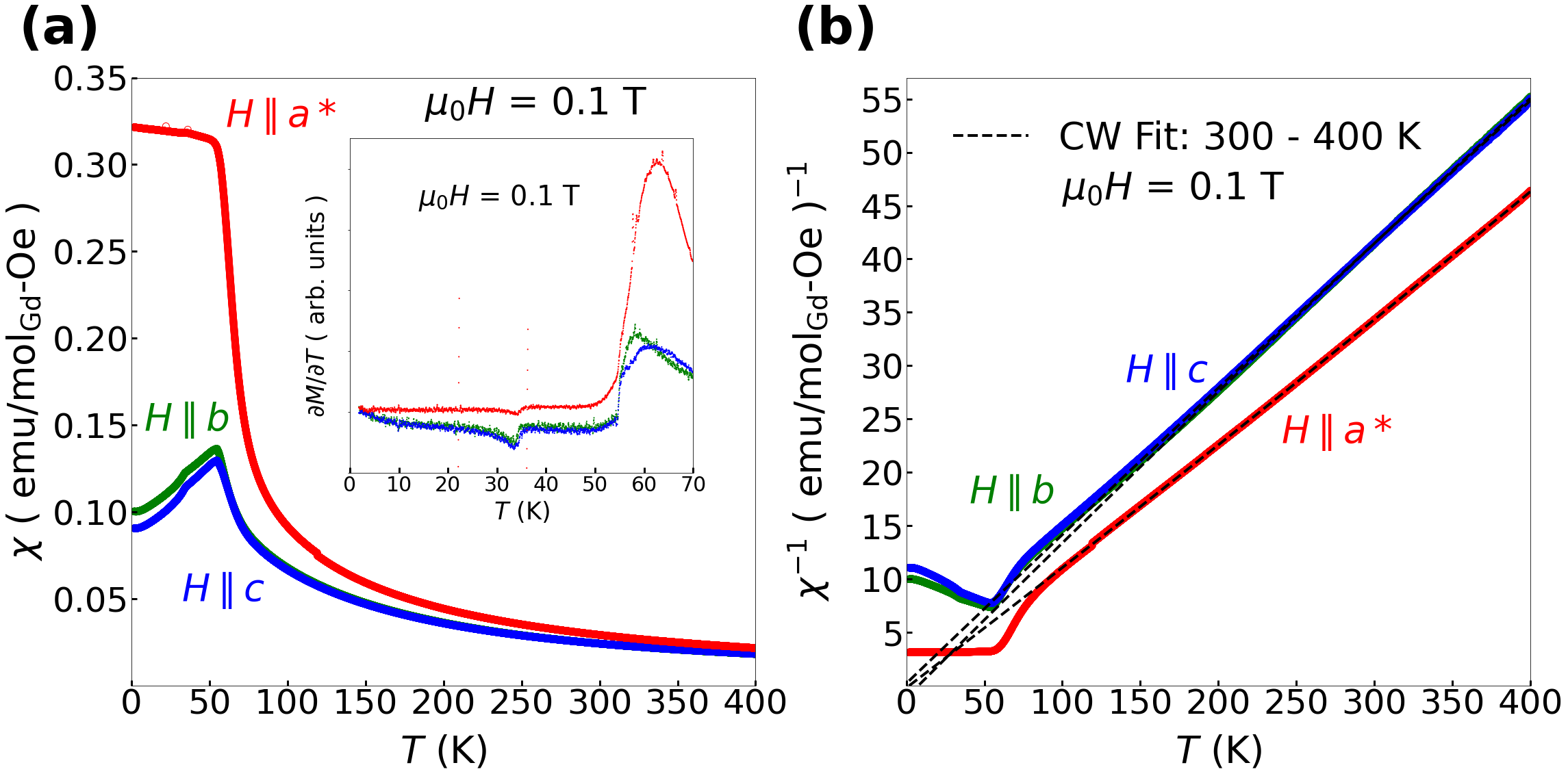}
	\caption{\textbf{(a)} Magnetic susceptibility plotted as a function of temperature in an applied field of $0.1 \, \text{T}$ along each of the \textit{a*}, \textit{b}, and \textit{c} axes. \textbf{(b)} Inverse magnetic susceptibility and Curie-Weiss fits (black, dashed lines) in an applied field of $0.1 \, \text{T}$ along each of the \textit{a*}, \textit{b}, and \textit{c} axes.}
	\label{fig:Gd2B5_CW}
\end{figure}

Figure \ref{fig:Gd2B5_CW}(a) shows the magnetic susceptibility of Gd\textsubscript{2}B\textsubscript{5} measured in an applied magnetic field of $0.1 \, \text{T}$ with field aligned parallel to the \textit{a*}, \textit{b}, and \textit{c} axes. The susceptibility exhibits a sharp downward kink at $T_1$ for field aligned with the \textit{b} and \textit{c} axes, but appears to level off at $T_1$ for field aligned with the \textit{a*} axis. The inset of \ref{fig:Gd2B5_CW}(a) shows the derivative $\partial \chi/\partial T$ plotted as a function of temperature, with clear features visible at $T_1$ and $T_2$ for all field orientations.

Figure \ref{fig:Gd2B5_CW}(b) shows the inverse magnetic susceptibility for the applied magnetic field oriented along the \textit{a*}, \textit{b}, and \textit{c} axes. The data was fit between $300-400 \, \text{K}$ (dashed, black lines) by a Curie-Weiss model with a constant offset via the following equation:

\begin{equation}
    \chi = \frac{1}{T-\theta_{CW}} \frac{N_A}{3k_B} \mu_{eff}^2 + \chi_0
\end{equation}

\noindent
where $\chi$ is the magnetic susceptibility per mole, $\theta_{CW}$ is the Curie temperature, $N_A$ is Avogadro's number, $\chi_0$ is the $T$-independent susceptibility component, and $\mu_{eff}$ is the effective moment in units of $\mu_B$. The parameters obtained from this fit are displayed in Table \ref{T: CW_Gd}. The free-space effective moment of Gd\textsuperscript{3+}  is $7.94 \, \mu_B$. The effective moments for field aligned with the \textit{b} and \textit{c} axes are similar and are both less than the free-space value, whereas the effective moment for field aligned with the \textit{a*} axis is greater than the free-space value.

\begin{table}[H]
\caption{\label{T: CW_Gd}Parameters from the Curie-Weiss fit of Gd\textsubscript{2}B\textsubscript{5} crystal X1 between $300-400 \, \text{K}$ in a field of $\mu_0 H = 0.1 \, \text{T}$ aligned along the \textit{a*}, \textit{b}, and \textit{c} axes of the crystal. The free-space effective moment of Gd\textsuperscript{3+}  is $7.94 \, \mu_B$.}
\centering{
\begin{tabular}{p{3.5cm} P{1.5cm} P{1.5cm} P{1.5cm}}
\toprule
  Parameter &  $H \parallel a^*$ &  $H \parallel b$ &  $H \parallel c$   \\ [0.8ex] 
  \hline
  $\theta_{CW}$ (K) & 2.02 & 7.81 & -1.34 \\ [0.6ex] 
  $\chi_0$ (emu/mol\textsubscript{Gd}-kOe) & -0.918 & 0.739 & 0.510 \\ [0.6ex] 
  $\mu_{eff}$ ($\mu_B$/Gd) & 8.46 & 7.54 & 7.39 \\ [0.6ex] 
\bottomrule
\end{tabular}
}
\end{table}

\begin{figure}[H]
	\centering 
	\includegraphics[width=1.0
    \linewidth]{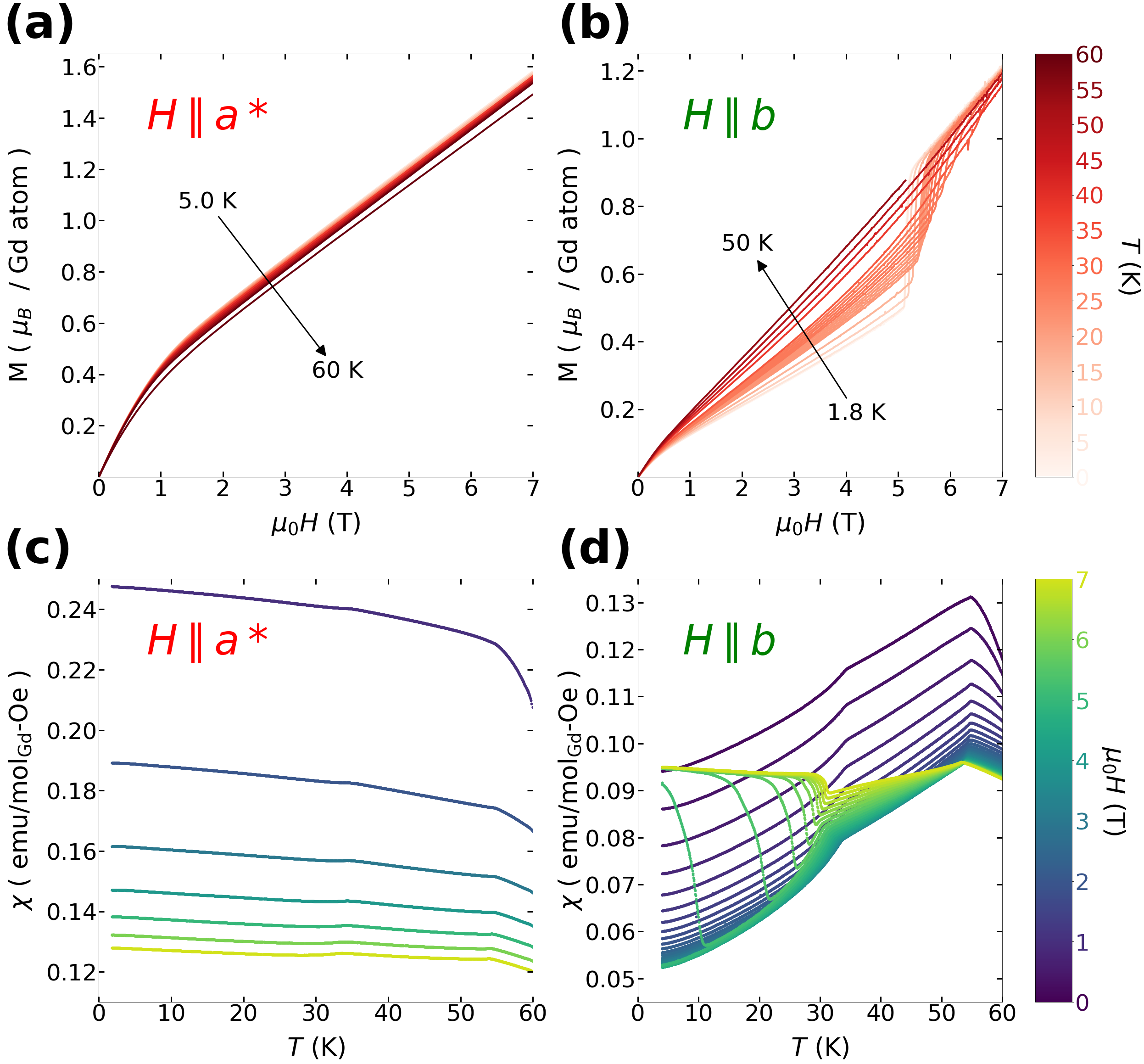}
	\caption{Magnetization plotted as a function of field at fixed temperatures up to $60\, \text{K}$ with the applied field aligned to the \textbf{(a)} \textit{a*} axis and \textbf{(b)} \textit{b} axis. Magnetic susceptibility plotted as a function of temperature in fixed applied magnetic fields up to $7\, \text{T}$ aligned with the \textbf{(c)} \textit{a*} axis and \textbf{(d)} \textit{b} axis.}
	\label{fig:Gd_MasterMag}
\end{figure}

Figures \ref{fig:Gd_MasterMag}(a-b) show the magnetization as a function of applied magnetic field along the \textit{a*} and \textit{b} axes. In contrast to the behavior when the field is aligned with the \textit{a*} axis, the magnetization exhibits a sharp jump upon entering the M$_\perp$ phase when the field is aligned with the \textit{b} axis. Figures \ref{fig:Gd_MasterMag}(c-d) show the magnetic susceptibility plotted as a function of temperature in a fixed applied magnetic field along the \textit{a*} and \textit{b} axes. The magnetic susceptibility exhibits kinks associated with the PM$\leftrightarrow$M\textsubscript{1} and M\textsubscript{1}$\leftrightarrow$M\textsubscript{2} phase transitions for all field orientations. For field orientations parallel to the \textit{b} axis, a sharp kink appears upon entering the M$_\perp$ phase. The behavior of the magnetization for field aligned with the \textit{c} axis (not shown here) was qualitatively the same as that for field aligned with the \textit{b} axis.

\begin{figure}[H]
	\centering 
	\includegraphics[width=0.95
    \linewidth]{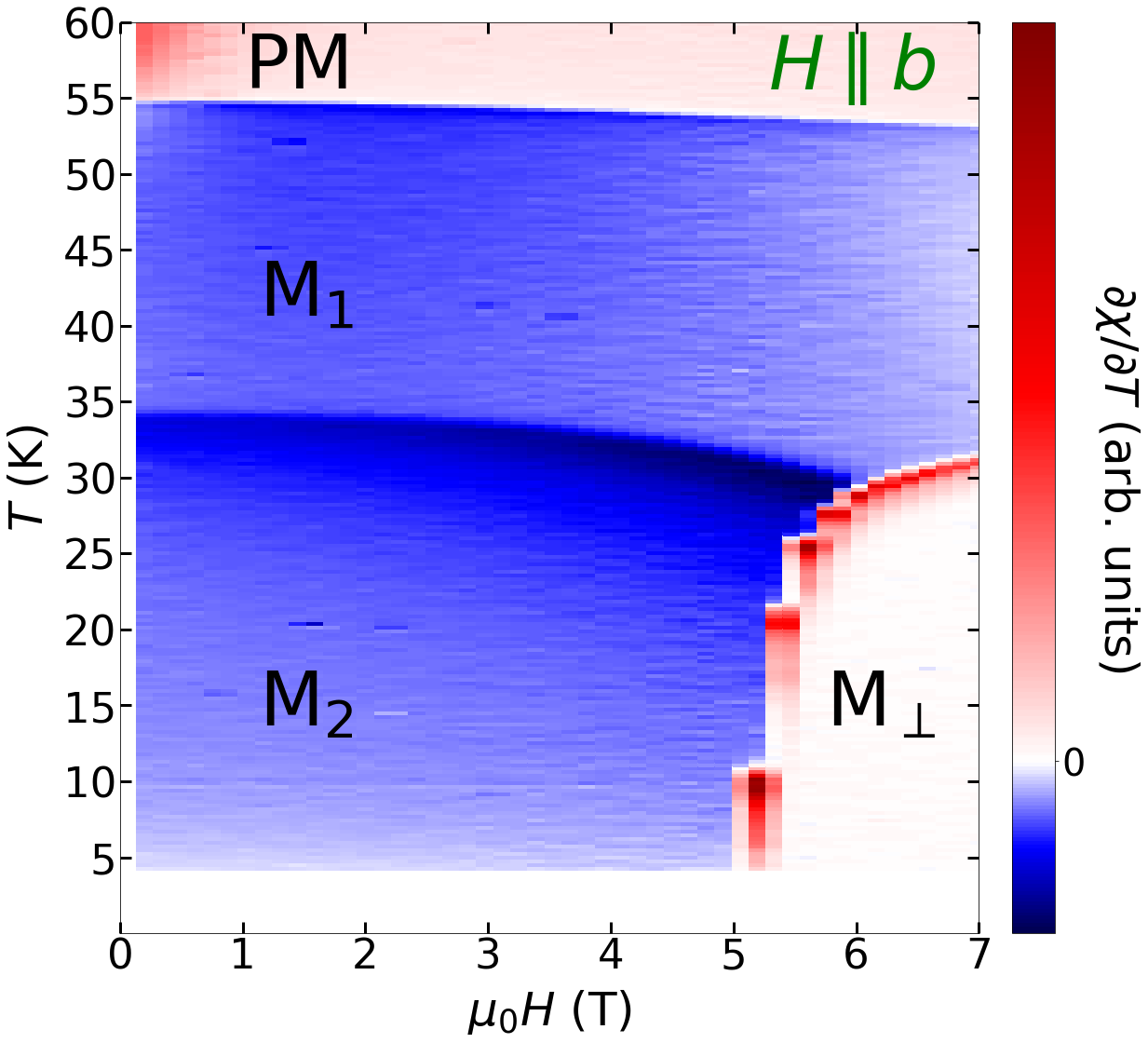}
	\caption{A colorplot created from measurements of the derivative of the magnetic susceptibility with respect to temperature for fixed magnetic fields applied along the \textit{b} axis. The PM, M\textsubscript{1}, M\textsubscript{2}, and M$_\perp$ phases are all labeled.}
	\label{fig:Gd_MvTb_CP}
\end{figure}

Figure \ref{fig:Gd_MvTb_CP} presents a colorplot generated from measurements of the derivative of the magnetic susceptibility with respect to temperature for fixed magnetic fields applied along the \textit{b} axis. This colorplot clearly shows the zero-field transitions into the M\textsubscript{1} and M\textsubscript{2} phases as well as the transition into the M$_\perp$ phase region in the bottom right corner of the plot. Note that the M$_\perp$ phase boundary intersects the M\textsubscript{1}$\leftrightarrow$M\textsubscript{2} phase boundary.

\subsection{Magnetic Torque}
\label{Magnetic Torque}

Magnetic torque was measured using NANOSENSORS cantilevers on single crystals of Gd\textsubscript{2}B\textsubscript{5}. One sample was measured in fields up to $14 \, \text{T}$ in a Quantum Design Physical Property Measurement System (PPMS), and a second sample was measured up to $55 \, \text{T}$ at the National High Magnetic Field Laboratory's Pulsed Field Facility at Los Alamos National Laboratory.

\begin{figure}[H]
	\centering 
	\includegraphics[width=1.0
    \linewidth]{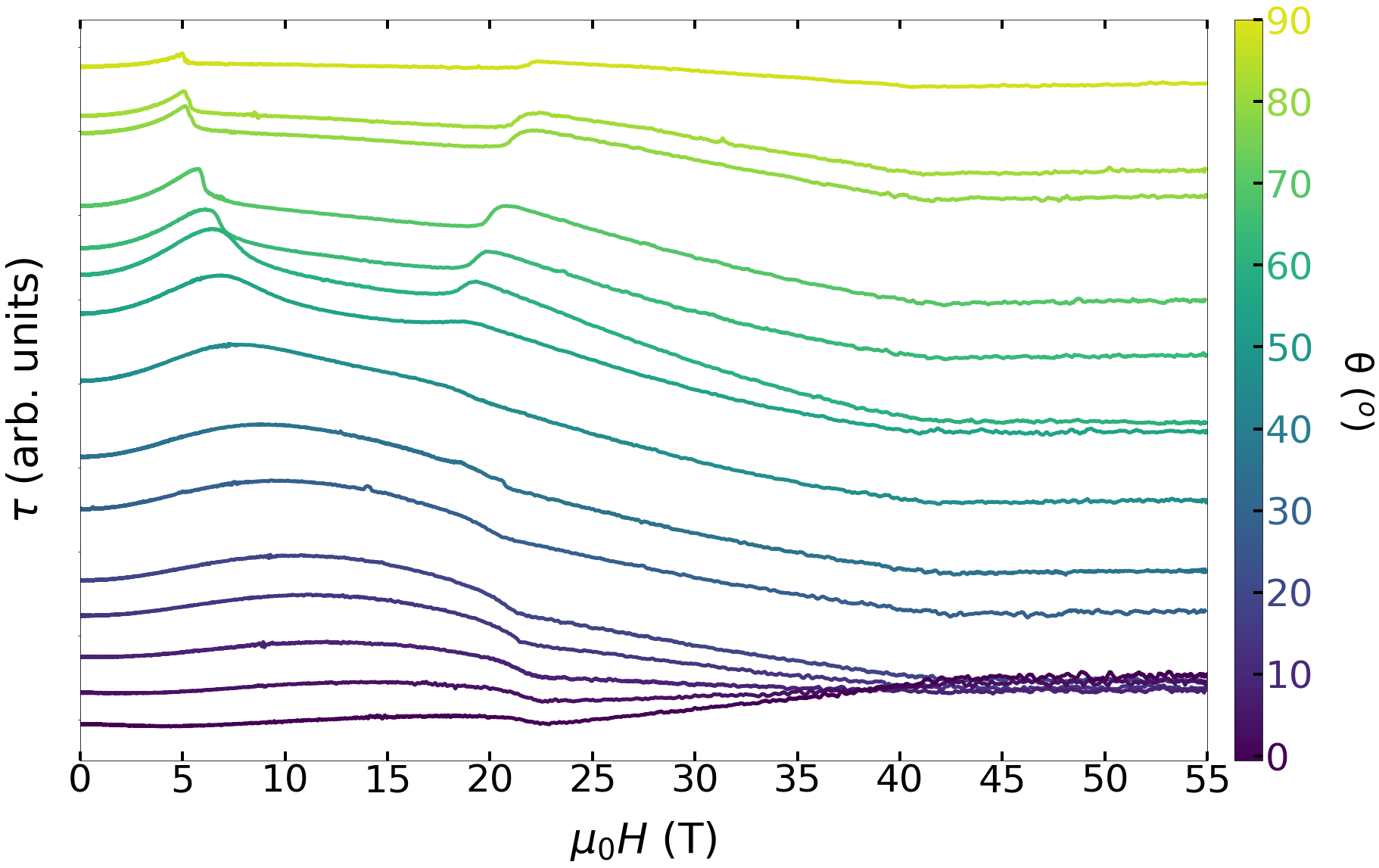}
	\caption{\textbf{(a)} Magnetic torque measured at fixed temperatures between $1.5$ and $1.6 \, \text{K}$ as a function of applied field for various out-of-plane field-angles $\theta$. Note that all of these data have been offset vertically for clarity.}
	\label{fig:TorOutP}
\end{figure}

For magnetic field applied close to parallel to the \textit{a*} axis (close to $\theta = 0$\textsuperscript{o}), the isothermal magnetic torque exhibits kink features concurrent with the PM$\leftrightarrow$M\textsubscript{1} and M\textsubscript{1}$\leftrightarrow$M\textsubscript{2} phase transitions. For magnetic field applied perpendicular to the \textit{a*} axis (close to $\theta = 90$\textsuperscript{o}), two features appear at $5 \, \text{T}$ and $22 \, \text{T}$ associated with entering and leaving the M$_\perp$ phase. Figure \ref{fig:TorOutP} presents torque data measured with field applied to an out-of-plane angle $\theta$ with respect to the \textit{a*} axis at fixed temperatures between $1.5$ and $1.6 \, \text{K}$. As the applied field direction is rotated out of the $(1 \, 0 \, 0)$ plane, the features associated with the M$_\perp$ phase are brought closer together in field before vanishing.

\subsection{Heat Capacity}
\label{Heat Capacity}

Heat capacity was measured using a Quantum Design PPMS system with the heat capacity module enabled. A single crystal $C1$ was affixed to heater platform using Apiezon N grease. A background heat capacity measurement was taken of the grease prior to mounting the crystal such that it could be subtracted from the total heat capacity after the crystal was mounted.

\begin{figure}[H]
	\centering 
	\includegraphics[width=1.0
    \linewidth]{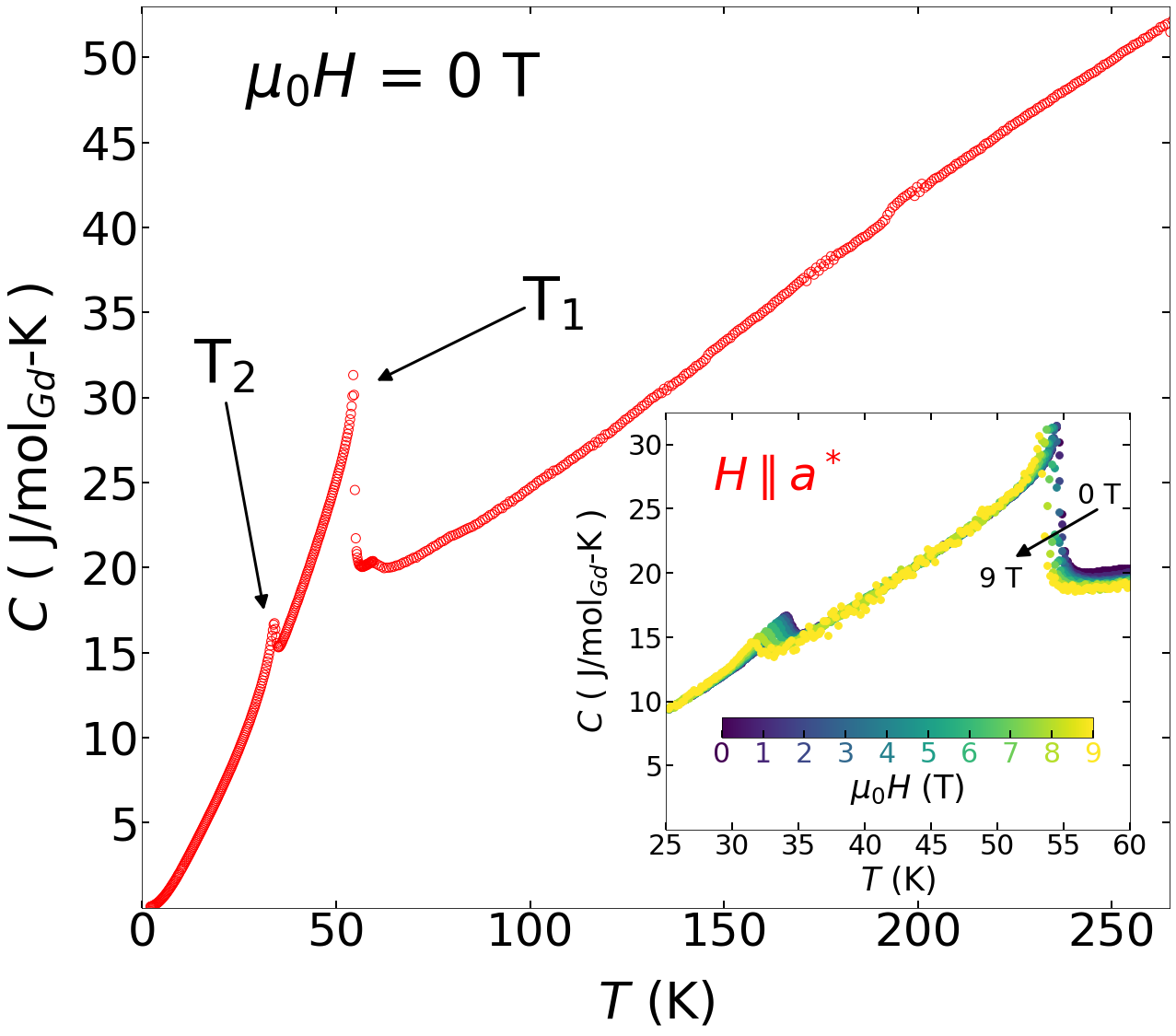}
	\caption{Zero-field heat capacity of crystal $C1$ with the two transition temperatures $T_1$ and $T_2$ labeled. The inset shows the heat capacity measured in magnetic fields up to $9 \, \text{T}$ applied parallel to the \textit{a*} axis.}
	\label{fig:HCT}
\end{figure}

Figure \ref{fig:HCT} shows the zero-field heat capacity of Gd\textsubscript{2}B\textsubscript{5} plotted under $260 \, \text{K}$ with the $T_1$ and $T_2$ transition temperatures labeled. The inset of this figure shows the behavior of the heat capacity in magnetic fields applied parallel to the \textit{a*} axis up to $9 \, \text{T}$. The behavior of the features as field increases is consistent with that of the PM$\leftrightarrow$M\textsubscript{1} and M\textsubscript{1}$\leftrightarrow$M\textsubscript{2} phase transitions seen in other probes.

\subsection{Phase Diagram for $H \parallel a^*$}
\label{Phase Diagram}

\begin{figure}[H]
	\centering 
	\includegraphics[width=1.0
    \linewidth]{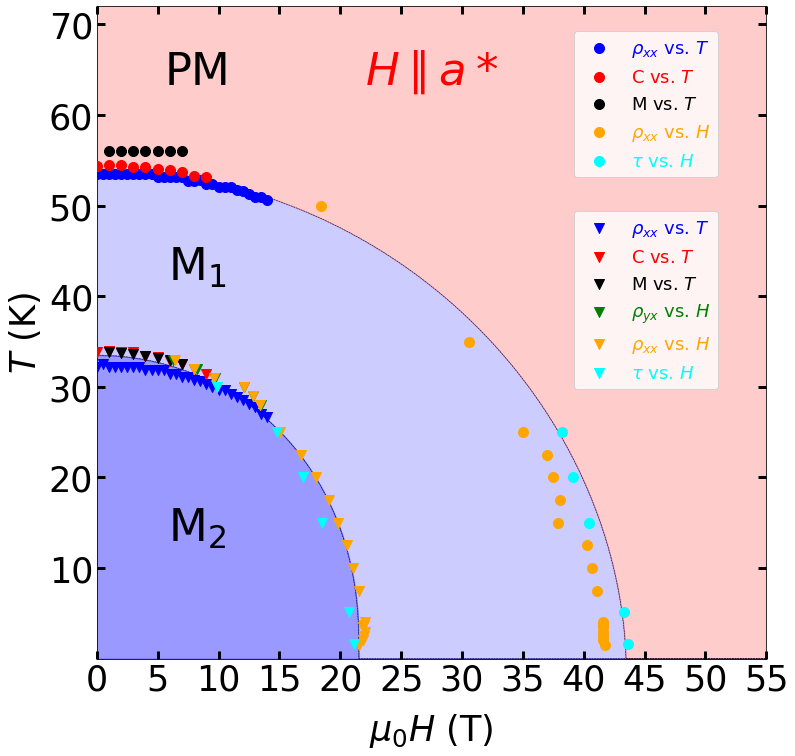}
	\caption{Phase diagram for Gd\textsubscript{2}B\textsubscript{5} with field parallel to the \textit{a*} axis showing the high temperature paramagnetic phase (PM) and the two low temperature magnetic phases (M\textsubscript{1} and M\textsubscript{2}). The black curves serve as a guide for the eye.}
	\label{fig:Gd2B5PD}
\end{figure}

The phase transitions observed in Gd\textsubscript{2}B\textsubscript{5} for field oriented parallel to the \textit{a*} axis can be mapped out into a phase diagram by tracking the corresponding features appearing in the measurements taken heretofore for this field orientation. This phase diagram is shown in Figure \ref{fig:Gd2B5PD} wherein circle-markers are used to delineate the PM$\leftrightarrow$M\textsubscript{1} boundary and triangle-markers are used to delineate the M\textsubscript{1}$\leftrightarrow$M\textsubscript{2} boundary. The markers are located at the positions $T^*$ of the features in longitudinal resistivity (blue marker), heat capacity (red marker), and magnetization (black marker) all measured in fixed field, as well as the positions $H^*$ of the feature in Hall resistivity (green marker), longitudinal resistivity (orange marker), and magnetic torque (cyan marker) measured at fixed temperature (see Appendix \ref{sec: Additional Phase Diagram Details} for additional details).

\section{Discussion}
\label{sec:Discussion}

The Gd\textsubscript{2}B\textsubscript{5} system realizes a highly anisotropic magnetic phase diagram exhibiting two zero-field phases M\textsubscript{1} and M\textsubscript{2} as well as a field-induced M$_\perp$ phase which appears when the applied magnetic field is close to pointing in the $(1 \, 0 \, 0 )$ plane (perpendicular to the \textit{a*} axis). The boron network layers in this plane interweave the three dimensional structure of gadolinium atoms and strongly effect the electronic properties of the system. The magnetic phases uncovered in  Gd\textsubscript{2}B\textsubscript{5} have been probed with magnetic fields applied both in and out of the $(1 \, 0 \, 0 )$ plane through both electrical transport and thermodynamic measurements. 

At high temperatures, the paramagnetic PM phase exhibits typical Curie-Weiss paramagnetic behavior for a free Gd\textsuperscript{3+} ion, before realizing a sharp kink in the magnetic susceptibility as the system is cooled into the M\textsubscript{1} phase. When field is applied parallel to the \textit{a*} axis the magnetic susceptibility levels off, whereas it kinks down sharply for field parallel to the \textit{b} and \textit{c} axes (see Fig. \ref{fig:Gd2B5_CW}). This behavior suggests that the magnetic moments in the M\textsubscript{1} phase point primarily in the $(1 \, 0 \, 0 )$ plane. The M\textsubscript{1}$\leftrightarrow$M\textsubscript{2} phase transition is more subtle as it appears as a slight kink in the magnetic susceptibility for all orientations. Aside from this kink, as the behavior of the magnetic susceptibility does not change significantly through this phase boundary, it is likely that the magnetic moments in the M\textsubscript{2} phase also point in the $(1 \, 0 \, 0 )$ plane. Both of the PM$\leftrightarrow$M\textsubscript{1} and M\textsubscript{1}$\leftrightarrow$M\textsubscript{2} phase transitions appear as features in electrical transport, magnetization, torque, and heat capacity measurements. From the torque data acquired at National High Magnetic
Field Laboratory’s Pulsed Field Facility at Los Alamos National Laboratory, at $1.5-1.6 \, \text{K}$ the M\textsubscript{2} phase appears to persist until about $21 \, \text{T}$ and the M\textsubscript{1} phase appears to persist until about $44 \, \text{T}$.

When a magnetic field is aligned with the \textit{a*} axis, The Hall resistivity exhibits a pronounced minimum at about $11.5 \, \text{T}$ whereafter SdH quantum oscillations appear in the M\textsubscript{2} phase. The minimum in the Hall resistivity may arise from a restructuring of the Fermi surface through a field-induced Lifshitz transition. New pockets in this restructured Fermi surface may then cause these oscillations to suddenly appear right after the transition. The M\textsubscript{2} phase exhibits two Fermi pockets denoted $\alpha_2$ and $\beta_2$. The $\alpha_2$ pocket appears at a frequency of $184 \, \text{T}$ and the $\beta_2$ pocket appears at a frequency of $443 \, \text{T}$. SdH oscillations in the M\textsubscript{1} phase exhibit three Fermi pockets denoted $\alpha_1$, $\beta_1$, and $\gamma_1$. In comparison to the the $\alpha_2$ and $\beta_2$ pockets in the M\textsubscript{2} phase, the $\alpha_1$ and $\beta_1$ pockets in the M\textsubscript{1} phase respectively appear at shifted frequencies and have reduced effective masses. This shift, as well as the appearance of an additional $\gamma_1$ pocket, suggest a change in the electronic band structure when the M\textsubscript{1}$\leftrightarrow$M\textsubscript{2} phase boundary is crossed.

When a magnetic field is applied in the $(1 \, 0 \, 0 )$ plane, the system enters the M$_\perp$ phase between $5-6 \, \text{T}$ depending on the azimuthal field-angle within the plane. The isothermal magnetization exhibits a sharp jump as the system enters the M$_\perp$ phase which suggests a possible metamagnetic transition. The longitudinal and transverse resistivities also exhibit a sharp feature as the system enters the M$_\perp$ phase which appears to be hysteretic. The transverse resistivity measured for this field orientation may be explained by the presence of an anomalous Hall effect associated with the M$_\perp$ phase or by the low symmetry of the monoclinic system allowing for off-diagonal antisymmetric components of the  resistivity tensor. From in-plane torque measurements, the M$_\perp$ phase appears to persist until about $22 \, \text{T}$. When rotating out of the $(1 \, 0 \, 0 )$ plane the M$_\perp$ phase appears to be stable within about $\pm 20$\textsuperscript{o}. For a narrow band of in-plane field-angles, a pocket of negative magnetoresistance appears above $12 \, \text{T}$ which could suggest the presence of a either a new in-plane phase of a highly anisotropic upper critical field for the M$_\perp$ phase.

\section{\label{sec:Conclusion} Conclusion}

The presented data show that the Gd\textsubscript{2}B\textsubscript{5} system exhibits a rich magnetic phase diagram with two zero-field phases M\textsubscript{1} and M\textsubscript{2} as well as a field-induced M$_\perp$ phase all probed through a combination of electrical transport and thermodynamic measurements. Notable additional features of the system include hysteresis in the resistivity when measured across the M$_\perp$ phase as a function of magnetic field, and the appearance of quantum oscillations above a minimum in the Hall resistivity at $11.5 \, \text{T}$ which may appear after a Lifshitz transition of the Fermi surface when the magnetic field is applied parallel to the \textit{a*} axis. 

An important future research direction would be to determine the nature of the magnetic order which yields the aforementioned phases. Ordinarily, a neutron diffraction experiment could be used to probe the magnetic structure; this is difficult in Gd\textsubscript{2}B\textsubscript{5} due to naturally abundant isotopes of gadolinium and boron both having high neutron absorption cross sections (neutron poisons) and thereby requiring the use of isotopically enriched gadolinium and boron to produce suitable crystals of Gd\textsubscript{2}B\textsubscript{5}. To overcome this, resonant elastic x-ray scattering (REXS) techniques will be valuable for probing the magnetic structure. Lorentz transmission electron microscopy (Lorentz TEM) would additionally provide a probe for the in-plane magnetic moment which would be important for studying the structure of the magnetic phases; Lorentz TEM would be especially valuable for probing potential skyrmions in the M$_\perp$ phase. A study of the electronic band structure of the system through angle-resolved photoemission spectroscopy (ARPES) may also be valuable to determine if the observed quantum oscillations in the M\textsubscript{2} phase actually reflect the zero-field band structure of the M\textsubscript{2} phase, or if the band structure was significantly restructured through a Lifshitz transition. Gd\textsubscript{2}B\textsubscript{5} provides an exciting opportunity for further study of a highly anisotropic system rich in magnetic phases which has heretofore not been extensively studied.

\begin{acknowledgments}
This work was funded, in part, by the Gordon and Betty Moore Foundation EPiQS Initiative, Grant No. GBMF9070 to J.G.C (instrumentation development) the Army Research Office, Grant No. W911NF-24-1-0234 (material characterization), and the Center for Advancement of Topological Semimetals, an Energy Frontier Research Center funded by the US Department of Energy (DOE), Office of Science, Basic Energy Sciences (BES), through the Ames Laboratory (contract no. DE-AC02-07CH11358) (pulsed-field experiments). A portion of this work was performed at the National High Magnetic Field Laboratory, which is supported by National Science Foundation Cooperative Agreement No. DMR-2128556 and the State of Florida. A.H.M. acknowledges the support by JSPS Overseas Research Fellowships. This work was additionally carried out in part through the use of MIT.nano's facilities.

\end{acknowledgments}

\appendix

\section{X-ray Diffraction}
\label{sec: X-ray Diffraction}

\begin{figure}[H]
	\centering 
	\includegraphics[width=0.8
    \linewidth]{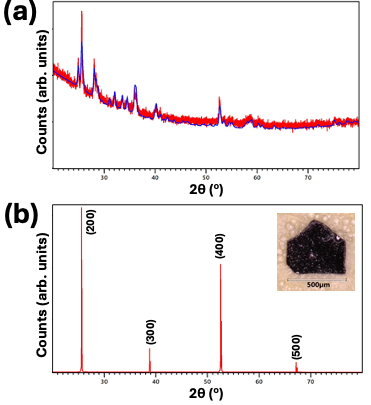}
	\caption{\textbf{(a)} Powder XRD spectrum for powdered single crystals of Gd\textsubscript{2}B\textsubscript{5} shown in red. A simulated diffraction pattern for Gd\textsubscript{2}B\textsubscript{5} is drawn in blue and fits the data well. \textbf{(b)} XRD spectrum for a $2\theta$ sweep of a single crystal laid flat on the sample stage. The peaks are indexed to multiples of $(1\,0\,0)$ as expected. The inset shows a Gd\textsubscript{2}B\textsubscript{5} single crystal with the $(1\,0\,0)$ plane parallel to the page.}
	\label{fig:Gd2S5_XRAY}
\end{figure}

The structure of the crystals obtained through the growth method outlined in Section \ref{sec:Growth Method} was determined through powder XRD. Figure \ref{fig:Gd2S5_XRAY}(a) shows a powder XRD spectrum obtained for powdered single crystals of Gd\textsubscript{2}B\textsubscript{5} (data shown in red). The blue curve is a simulated diffraction spectrum for Gd\textsubscript{2}B\textsubscript{5} which shows excellent agreement with the data. The single crystals of Gd\textsubscript{2}B\textsubscript{5} grew in a plate-like morphology with the plates parallel to the $(1\,0\,0)$ plane. Figure \ref{fig:Gd2S5_XRAY}(b) shows a $2\theta$ scan of a single crystal of Gd\textsubscript{2}B\textsubscript{5} laid flat on the sample stage such that the $(1\,0\,0)$ plane is parallel to the stage. The Bragg peaks in the diffraction spectrum were perfectly indexed to multiples of $(1\,0\,0)$ as expected. The orientation of the crystals was also confirmed using Laue XRD.

\section{SdH Quantum Oscillations}
\label{sec: SdH Quantum Oscillations}

SdH quantum oscillations were analyzed for both the M\textsubscript{2} and M\textsubscript{1} phases. The analysis starts by determining a window of magnetic field values $(H_1, H_2)$ that enclose a region of the resistivity exhibiting quantum oscillations and subtracting a smooth, polynomial background to isolate the quantum oscillations. As the quantum oscillations are periodic in $1/H$, these data may be transformed to be a function of $1/H$ and then Fourier transformed to isolate oscillations originating from different Fermi pockets. 

At fixed magnetic field, the temperature dependence of the amplitude of the first harmonic of an SdH quantum oscillation follows the Lifshitz-Kosevich formula given by:

\begin{equation}\label{RT}
    R_T = \frac{\pi \lambda}{\sinh{\pi \lambda}} \qquad , \qquad \lambda = \frac{2 \pi k_B T}{\beta H}
\end{equation}

\noindent
where $H$ is the magnetic field, $T$ is the temperature, and $\beta = e \hbar / mc$ ($m$ is the cyclotron mass) \cite{Shoenberg_1984}. If $H$ is expressed in units of Tesla and $T$ in units of Kelvin, then it can be shown that:

\begin{equation}
    \pi \lambda = (14.7) \, \tilde{m} \, \frac{T}{H} \qquad , \qquad \tilde{m} \equiv \frac{m}{m_e}
\end{equation}

\noindent
where $m_e$ is the free electron mass \cite{Shoenberg_1984}. To extract the effective mass $\tilde{m}$, the amplitudes of the peaks present in the Fourier transformed data taken at several different temperatures can be fit to the Lifshitz-Kosevich formula as a function of $T/H$ where $H$ is the fixed magnetic field at which the quantum oscillations are analyzed. Typically, this magnetic field $H$ is extracted in terms of the magnetic fields $H_1$ and $H_2$ as follows:

\begin{equation}
    H = \big(H_1^{-1} + H_2^{-1} \big)^{-1}
\end{equation}

\begin{figure}[H]
	\centering 
	\includegraphics[width=1.0
    \linewidth]{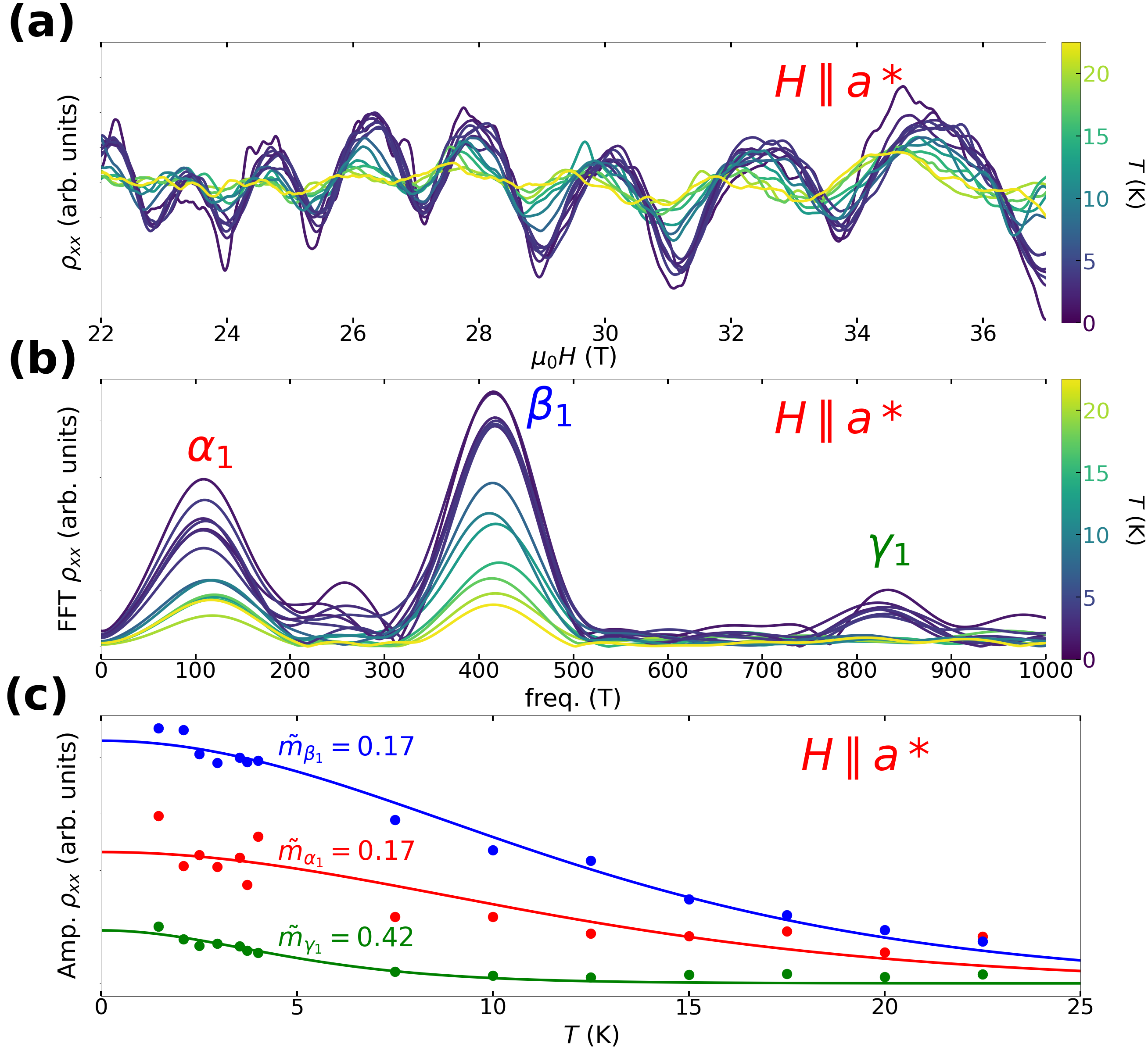}
	\caption{\textbf{(a)} Background subtracted longitudinal resistivity of device $R1$ in the M\textsubscript{1} phase showing SdH quantum oscillations. \textbf{(b)} Fourier transform of these data showing three pockets: $\alpha_1$, $\beta_1$, and $\gamma_1$. \textbf{(c)} Lifshitz-Kosevich fits of the amplitudes of each pocket with the extracted masses $\tilde{m} \equiv m^* / m_e$ shown.}
	\label{fig:QO_LANL}
\end{figure}

Analysis of the SdH oscillations in the M\textsubscript{1} phase is presented in Fig. \ref{fig:QO_LANL} for the longitudinal resistivity channel. These oscillations were all analyzed in the window between $22$ and $37 \, \text{T}$, and a quartic polynomial was fit to the data to serve as a background. Figure \ref{fig:QO_LANL}(a) shows the background subtracted longitudinal resistivity data plotted as a function of field for fixed temperatures between $1.46$ and $22.5 \, \text{K}$. Figure \ref{fig:QO_LANL}(b) shows the Fourier transforms of these data which reveal three pockets labeled $\alpha_1$, $\beta_1$, and $\gamma_1$ in the M\textsubscript{1} phase which contribute to the quantum oscillations. Figure \ref{fig:QO_LANL}(c) shows the Lifshitz-Kosevich fits of the amplitudes of these three pockets with the values of the extracted effective masses $\tilde{m}_{\alpha_1}$, $\tilde{m}_{\beta_1}$, and $\tilde{m}_{\gamma_1}$ labeled.

\begin{figure}[H]
	\centering 
	\includegraphics[width=1.0
    \linewidth]{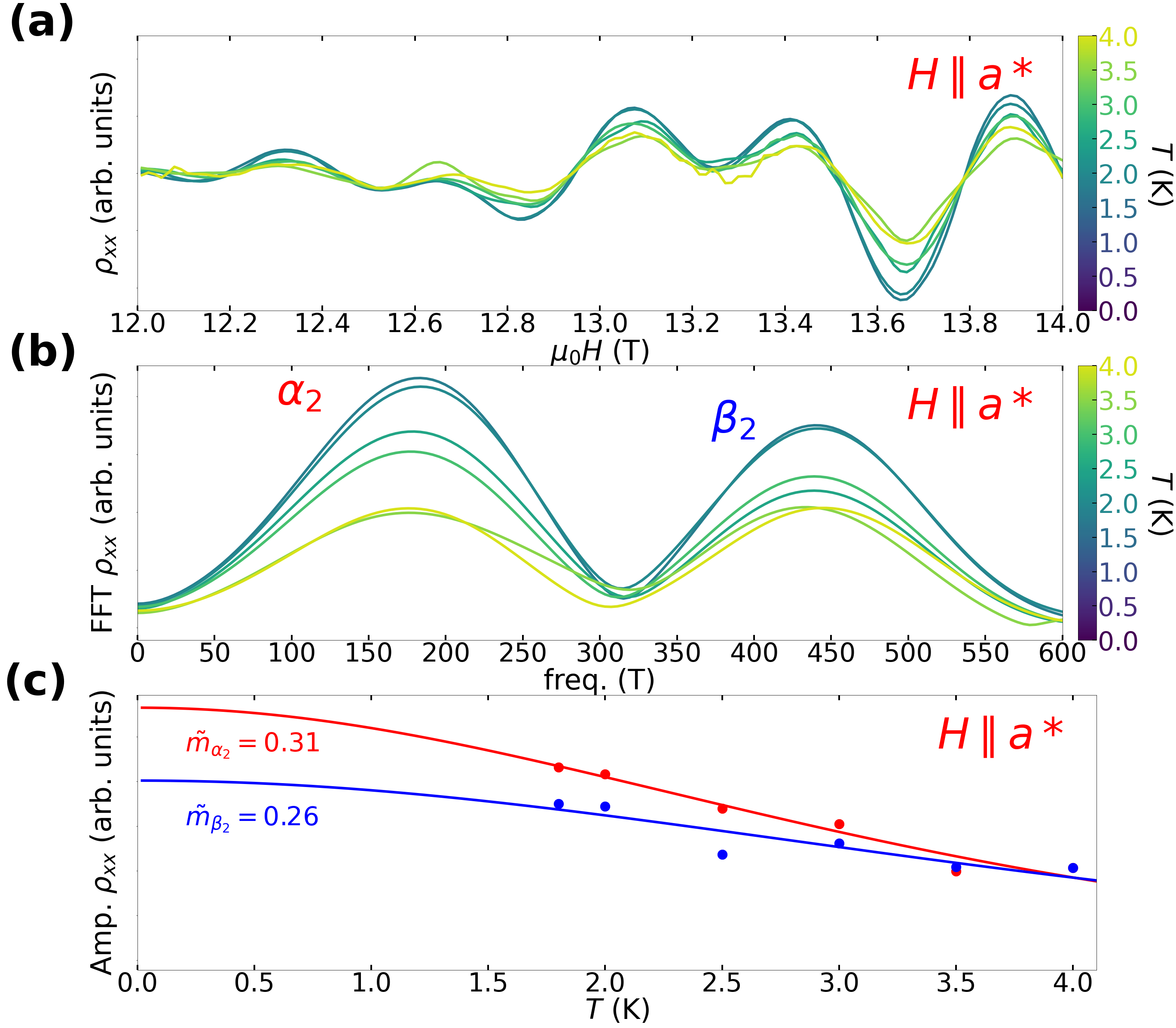}
	\caption{\textbf{(a)} Background subtracted longitudinal resistivity of device $R1$ in the M\textsubscript{2} phase showing SdH quantum oscillations. \textbf{(b)} Fourier transforms of these data showing two pockets: $\alpha_2$ and $\beta_2$. \textbf{(c)} Lifshitz-Kosevich fits of the amplitudes of each pocket with the extracted masses $\tilde{m} \equiv m^* / m_e$ shown.}
	\label{fig:M2QOLong}
\end{figure}

Analysis of the SdH oscillations in the M\textsubscript{2} phase is presented in Fig. \ref{fig:M2QOLong} for the longitudinal resistivity channel. These oscillations were all analyzed in the field window between $12$ and $14 \, \text{T}$, and a quadratic polynomial was fit to the data to serve as a background. Figure \ref{fig:M2QOLong}(a) shows the background subtracted longitudinal resistivity plotted as a function of field for fixed temperatures between $1.8$ and $4.0 \, \text{K}$. Figure \ref{fig:M2QOLong}(b) shows the Fourier transforms of these data which reveal two pockets labeled $\alpha_2$ and $\beta_2$ in the M\textsubscript{2} phase which contribute to the quantum oscillations. Figure \ref{fig:M2QOLong}(c) shows the Lifshitz-Kosevich fits of the amplitudes of these two pockets as a function of temperature with the values of the extracted effective masses $\tilde{m}_{\alpha_2}$ and $\tilde{m}_{\beta_2}$ labeled.

\section{Hysteresis in Electrical Transport}
\label{sec: Hysteresis in Electrical Transport}

Measurements of both longitudinal and transverse resistivity both exhibit hysteresis in association with the M$_\perp$ phase. To deal with this properly, both a downsweep $\rho^d(H)$ and an upsweep $\rho^u(H)$ must be measured such that the antisymmetrized transverse component may be extracted as follows:

\begin{gather}
    \rho_{yx}^d(H) = \frac12 \big( \rho^d(H) - \rho^u(-H)  \big)\\
    \rho_{yx}^u(H) = \frac12 \big( \rho^u(H) - \rho^d(-H)  \big)
\end{gather}

\noindent
A similar treatment may be applied to the symmetrized longitudinal channel to yield:

\begin{gather}
    \rho_{xx}^d(H) = \frac12 \big( \rho^d(H) + \rho^u(-H)  \big)\\
    \rho_{xx}^u(H) = \frac12 \big( \rho^u(H) + \rho^d(-H)  \big)
\end{gather}

\begin{figure}[H]
	\centering 
	\includegraphics[width=1.0
    \linewidth]{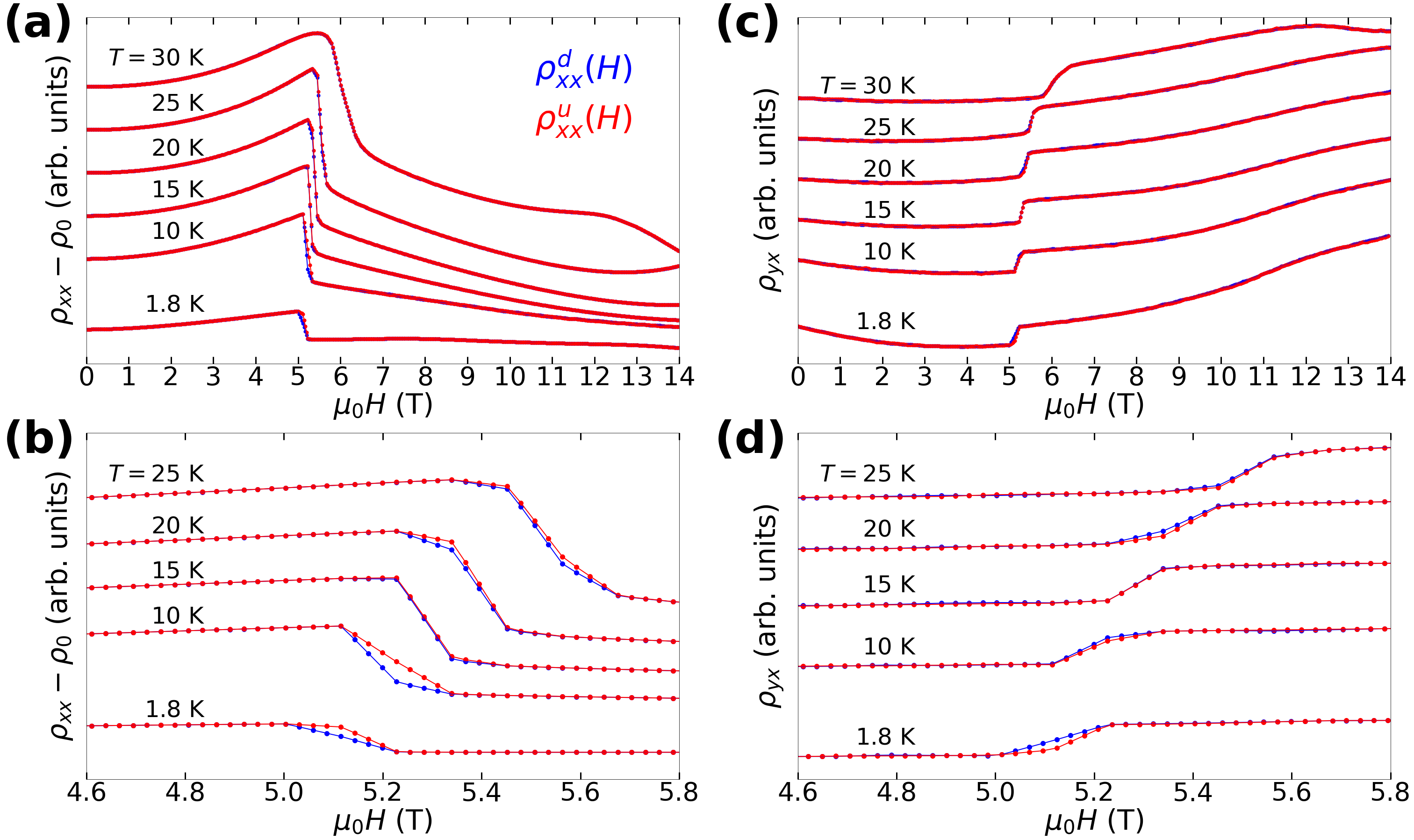}
	\caption{Longitudinal \textbf{(a-b)} and transverse \textbf{(c-d)} resistivity of device $R1$ at for an in-plane field angle ($\theta=90$\textsuperscript{o}) plotted as a function of applied field at several fixed temperatures. The blue traces show $\rho_{yx}^d(H)$ and $\rho_{xx}^d(H)$ whereas the red traces show $\rho_{yx}^u(H)$ and $\rho_{xx}^u(H)$. Note that all of these data have been offset vertically for clarity.}
	\label{fig:hyst}
\end{figure}

Figure \ref{fig:hyst} shows the longitudinal and transverse resistivity of device $R1$ symmetrized/antisymmetrized as described above at several fixed temperatures for magnetic field sweeps across the M$_\perp$ phase. These measurements were taken with the magnetic field applied in the $(1 \, 0 \, 0)$ plane ($\theta=90$\textsuperscript{o}) at an arbitrary azimuthal angle $\phi$. These data show a hysteretic feature in both channels up to $30 \, \text{K}$. Note that no hysteresis is observed in any other electrical transport measurements at field angles that do not cross the M$_\perp$ phase.

\section{Additional Phase Diagram Details}
\label{sec: Additional Phase Diagram Details}

The phase diagram shown in Figure \ref{fig:Gd2B5PD} delineates the PM$\leftrightarrow$M\textsubscript{1} and M\textsubscript{1}$\leftrightarrow$M\textsubscript{2} phase boundaries by marking the positions $T^*$ and $H^*$ of probes measured at fixed field and fixed temperature respectively. For both phase boundaries, the blue markers were obtained from the positions $T^*$ of extrema in the second derivative of longitudinal resistivity with respect to temperature (see Fig. \ref{fig:Res00}(b)), the red markers from positions $T^*$ of the peaks in heat capacity (see inset of Fig. \ref{fig:HCT}), and the black markers from positions $T^*$ of extrema in the first derivative of magnetization with respect to temperature (see the inset of Fig. \ref{fig:Gd2B5_CW}(a) and Fig. \ref{fig:Gd_MasterMag}(c)). 

\begin{figure}[H]
	\centering 
	\includegraphics[width=1.0
    \linewidth]{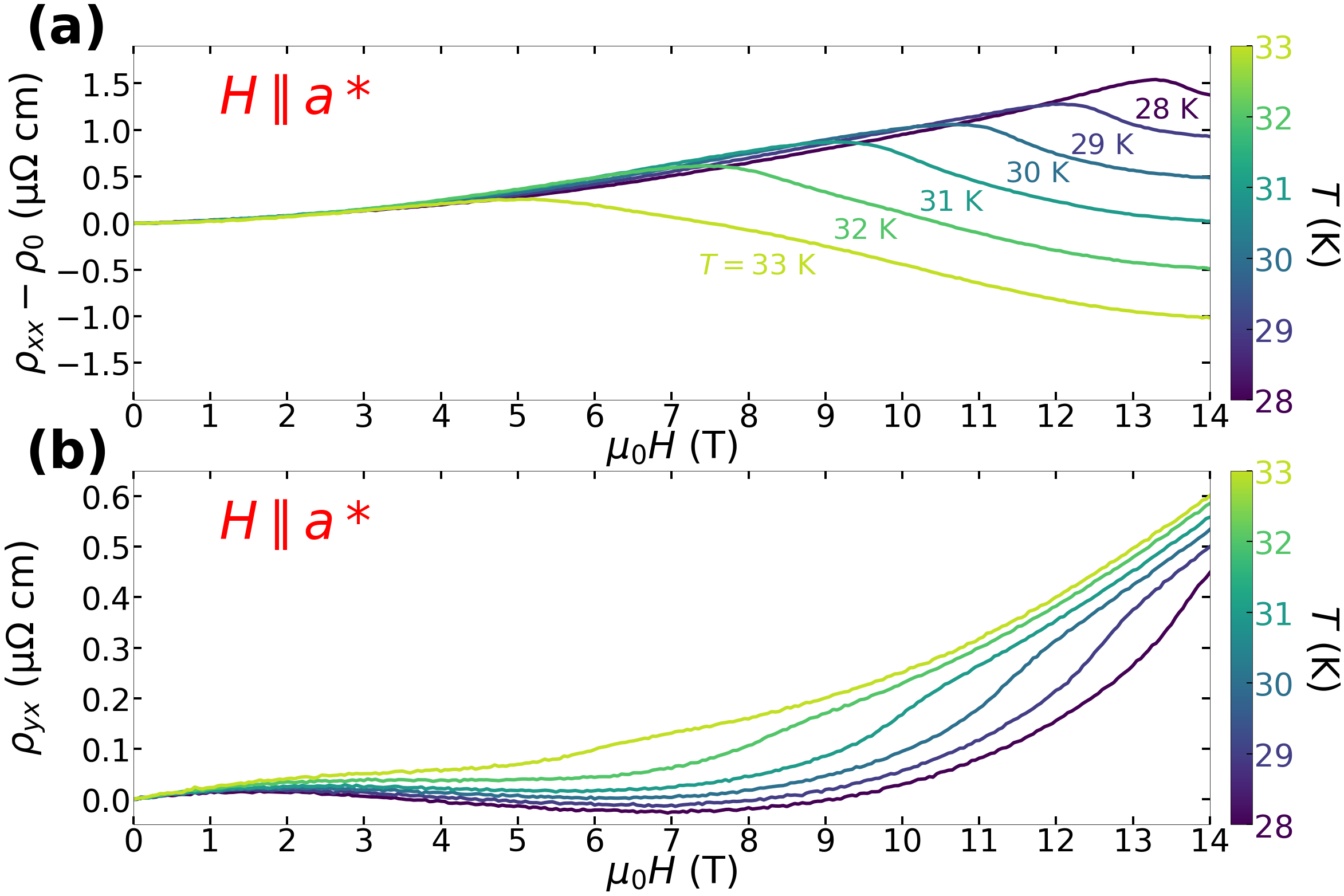}
	\caption{Longitudinal \textbf{(a)} and Hall \textbf{(b)} resistivity of device $R2$ in the vicinity of the M\textsubscript{1}$\leftrightarrow$M\textsubscript{2} phase boundary with field applied parallel to the \textit{a*} axis. The temperatures corresponding to the line colors in both plots are the same.}
	\label{fig:HsweepGd}
\end{figure}

Figure \ref{fig:HsweepGd} shows the longitudinal and Hall resistivity plotted as a function of magnetic field applied parallel to the \textit{a*} axis. When the M\textsubscript{1}$\leftrightarrow$M\textsubscript{2} phase boundary is crossed, both the longitudinal and Hall resistivities exhibits a kink. The M\textsubscript{1}$\leftrightarrow$M\textsubscript{2} phase boundary additionally has orange markers from the positions $H^*$ of extrema in the longitudinal resistivity and green markers from the positions $H^*$ of the first derivative of Hall resistivity with respect to applied field. The PM$\leftrightarrow$M\textsubscript{1} boundary additionally has orange markers from the positions $T^*$ of extrema in the second derivative of longitudinal resistivity (see Fig. \ref{fig:HsweepLANL}).

\begin{figure}[H]
	\centering 
	\includegraphics[width=1.0
    \linewidth]{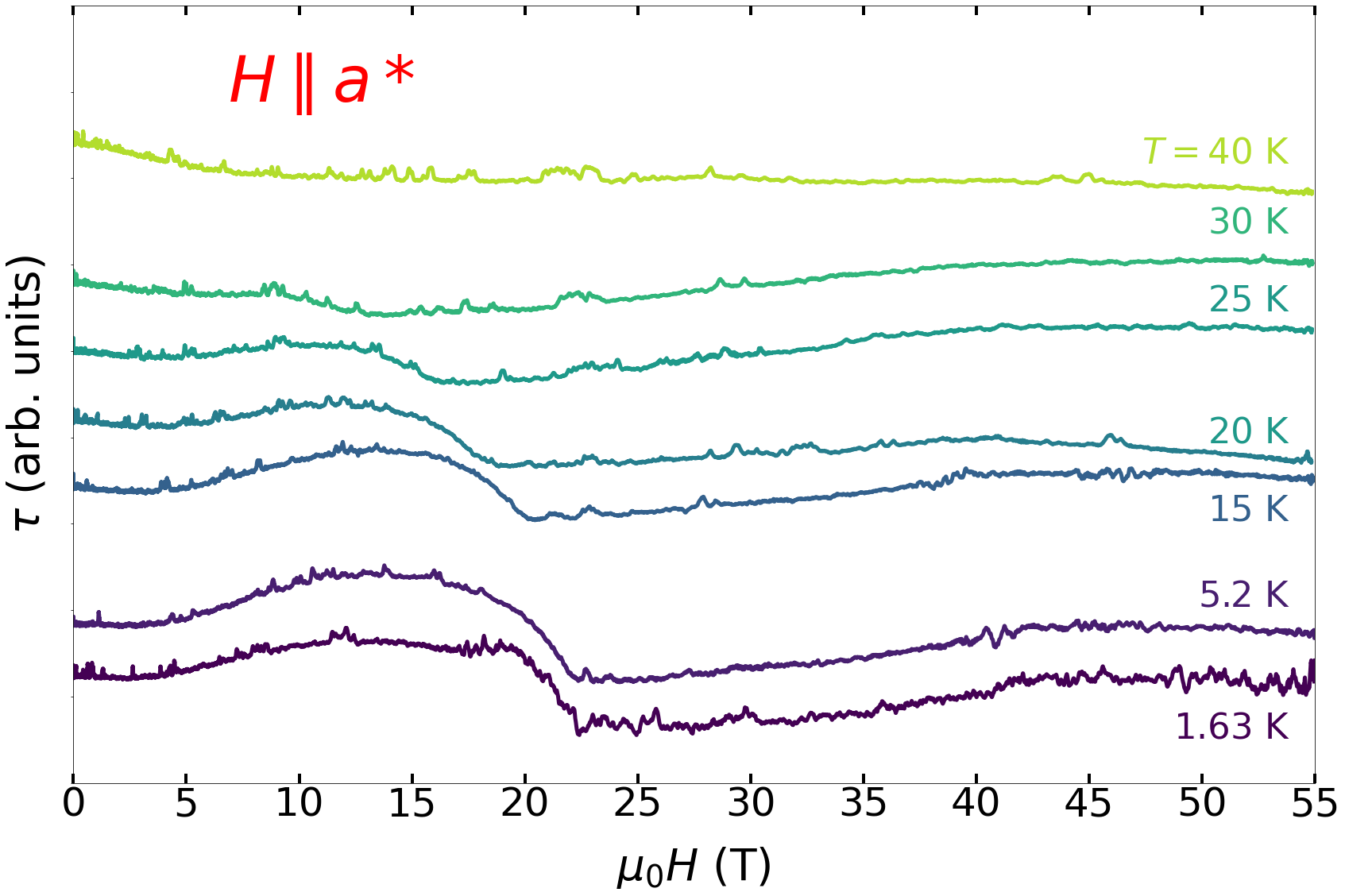}
	\caption{Magnetic torque measured at fixed temperatures for magnetic fields up to $55\, \text{T}$ applied parallel to the \textit{a*} axis. The two kinks in these curves arise from phase transitions out of the M\textsubscript{2} and M\textsubscript{1} phases. Note that all of these data have been offset vertically for clarity.}
	\label{fig:TorTemp}
\end{figure}

Figure \ref{fig:TorTemp} shows the isothermal magnetic torque plotted as a function of magnetic field applied parallel to the \textit{a*} axis at several fixed temperatures. The PM$\leftrightarrow$M\textsubscript{1} and M\textsubscript{1}$\leftrightarrow$M\textsubscript{2} phase boundaries are respectively delineated using cyan markers from the positions $H^*$ of the second and first derivatives of magnetic torque with respect to field.

%\nocite{*}

\clearpage
\bibliography{apssamp}% Produces the bibliography via BibTeX.

\end{document}